\documentclass{optica-article}
\journal{opticajournal} 
\articletype{Research Article}

\usepackage{graphicx}
\usepackage{bm}
\usepackage{caption}
\usepackage{subcaption}

\usepackage{ragged2e}
\DeclareCaptionJustification{justified}{\justifying}
\usepackage{diagbox}
\usepackage{braket,tensor,siunitx}
\usepackage{comment}

\DeclareMathOperator{\tr}{tr}
\DeclareMathOperator{\diag}{diag}
\graphicspath{{figures/}}
\allowdisplaybreaks
\usepackage{ulem}
\usepackage{lineno}

\begin{document}

\title{How to find optimal quantum states for optical micromanipulation and metrology in complex scattering problems: tutorial}


\author{Lukas M.\ Rachbauer,\authormark{1} Dorian Bouchet,\authormark{2} Ulf Leonhardt,\authormark{3} and Stefan Rotter\authormark{1,*}}
\address{\authormark{1}Institute for Theoretical Physics, Vienna University of Technology (TU Wien), A–1040 Vienna, Austria\\
\authormark{2}Universit{\'e} Grenoble Alpes, CNRS, LIPhy, 38000 Grenoble, France\\
\authormark{3}Department of Physics of Complex Systems, Weizmann Institute of Science, Rehovot 761001, Israel}
\email{\authormark{*}stefan.rotter@tuwien.ac.at}



\begin{abstract*}
    The interaction of quantum light with matter is of great importance to a wide range of scientific disciplines, ranging from optomechanics to high precision measurements. A central issue we discuss here, is how to make optimal use of both the spatial and the quantum degrees of freedom of light for characterizing and manipulating arbitrary observable parameters in a linear scattering system into which suitably engineered light fields are injected. Here, we discuss a comprehensive framework based on a quantum operator that can be assembled solely from the scattering matrix of a system and its dependence on the corresponding local parameter, making this operator experimentally measurable from the far-field using only classical light. From this, the effect of quantum light in the near-field, i.e., in the vicinity of the target object, can be inferred. Based on this framework, it is straightforward to formulate optimal protocols on how to jointly design both the spatial shape and the quantum characteristics of light for micromanipulation as well as for parameter estimation in arbitrarily complex media. Also the forces of the quantum vacuum naturally emerge from this formalism. The aim of our tutorial is to bring different perspectives into alignment and thereby build a bridge between the different communities of wave control, quantum optics, micromanipulation, quantum metrology and vacuum physics.
\end{abstract*}




\section{Introduction\label{sec:Introduction}}

The history of optics is marked by innovations that expanded our ability to manipulate light fields to make them useful for applications. Two research directions in which significant progress is currently being made in the creation of customized light fields are the domain of wavefront shaping, on the one hand, and the domain of quantum state engineering, on the other hand. While wavefront shaping is primarily concerned with the spatial patterns of light waves and their control, quantum state engineering deals with the quantum character of light and how to make it exploitable in practice. 

Until very recently, the developments in these two research directions have been largely disconnected from each other. This separation is all the more surprising as the areas of applications, that both wavefront shaping and the engineering of quantum states are concerned with, overlap significantly. Consider here, e.g., the field of imaging where considerable progress has recently been made, both with spatially shaped light fields \cite{Mosk2012,RotterGigan2017} and by engineering its quantum nature \cite{Moreau2019,FabreTreps2020}. In particular, by spatially shaping an incoming light beam it becomes possible to extract useful information from deeper layers of complex media \cite{Yoon2020} or to view across them \cite{Popoff2010_ImgTrans}. Quantum states of light, on the other hand, have not only lead to improvements in image resolution beyond the classical limit \cite{Kolobov2007}, but have also enabled entirely new imaging protocols, like ``ghost imaging'' \cite{Pittman1995,Padgett2017}. Take as another example, the field of optical micromanipulation, where the spatial engineering of light beams has yielded the versatile optical tweezers \cite{Ashkin1986,Phillips2014}, while quantum optomechanics has meanwhile achieved to cool down macroscopic objects into their motional ground states \cite{Chan2011,Delic2020,GonzalezBallestero2021} or to obtain squeezed light from micromechanical resonators \cite{SafaviNaeini2013}. Finally, also in the field of metrology, we now already understand very well how both the spatial \cite{Bouchet2020,Bouchet2021} and the quantum parameters \cite{Treps2003,Delaubert2006,Pinel2012,Taylor2013,Steinlechner2018} of light fields need to be organized in themselves to achieve extreme sensitivity in precision measurements.

Very recently, several works have appeared with the clear intention of bridging the gap between the exciting advances that have been made in the wavefront shaping community with those in quantum state engineering. Notable results are here, e.g., the precompensation of multiple photon scattering in complex media \cite{Defienne2016,Wolterink2016}, the spatial modulation of entangled photon pairs for tailoring high-dimensional quantum entanglement \cite{Defienne2018} and the combination of the phase sensitivity of NOON states with the orbital angular momentum of photons \cite{Hiekkamaeki2021_angular}. The controlled propagation of single-photon states through complex media \cite{Carpenter2013,Huisman2014,Defienne2014} has meanwhile also been extended to the programmable propagation of two-photon states through multi-mode fibers for advanced quantum information processing \cite{Defienne2016,Leedumrongwatthanakun2020}. Further results in this vein include the distillation of quantum images \cite{Defienne2019}, the real-time shaping of entangled photons using classical light \cite{Lib2020} and the successful unscrambling of entanglement through a complex medium \cite{Valencia2020}.

In this tutorial we identify and match concepts that are central both to wave control in complex media and to quantum metrology. For the former, this concept is known as the generalized Wigner-Smith operator \cite{Ambichl2017_Focusing,Horodynski2020}, whereas for the latter, the relevant concept is the generator of parameter translations \cite{Pang2017,Fiderer2019}. In the unified picture we provide here, we portray these two tools as mutually beneficial sides of a single coin. Correspondingly, we hope that the wave front shaping community will find this tutorial useful for the information it provides on how to include the quantum parameters of light in their protocols. The quantum metrology community, on the other hand, may find it useful to find hints on how to transfer and apply their knowledge to multi-mode complex scattering systems.

\section{Scattering theory\label{sec:Scat_theory}}

In the following Subsection \ref{subsec:Scat_cl_optics}, we provide an introduction to the scattering matrix formalism in classical optics. After outlining the fundamentals of multi-mode quantum optics and laying out the notation convention in Subsection \ref{subsec:Fund_quant_optics}, we describe how the classical scattering matrix determines the evolution of multi-mode quantum states of light in Subsection \ref{subsec:Scat_qu_optics}.

\subsection{Scattering in classical optics\label{subsec:Scat_cl_optics}}

\begin{figure*}[t]
    \centering
    \includegraphics[width=\textwidth]{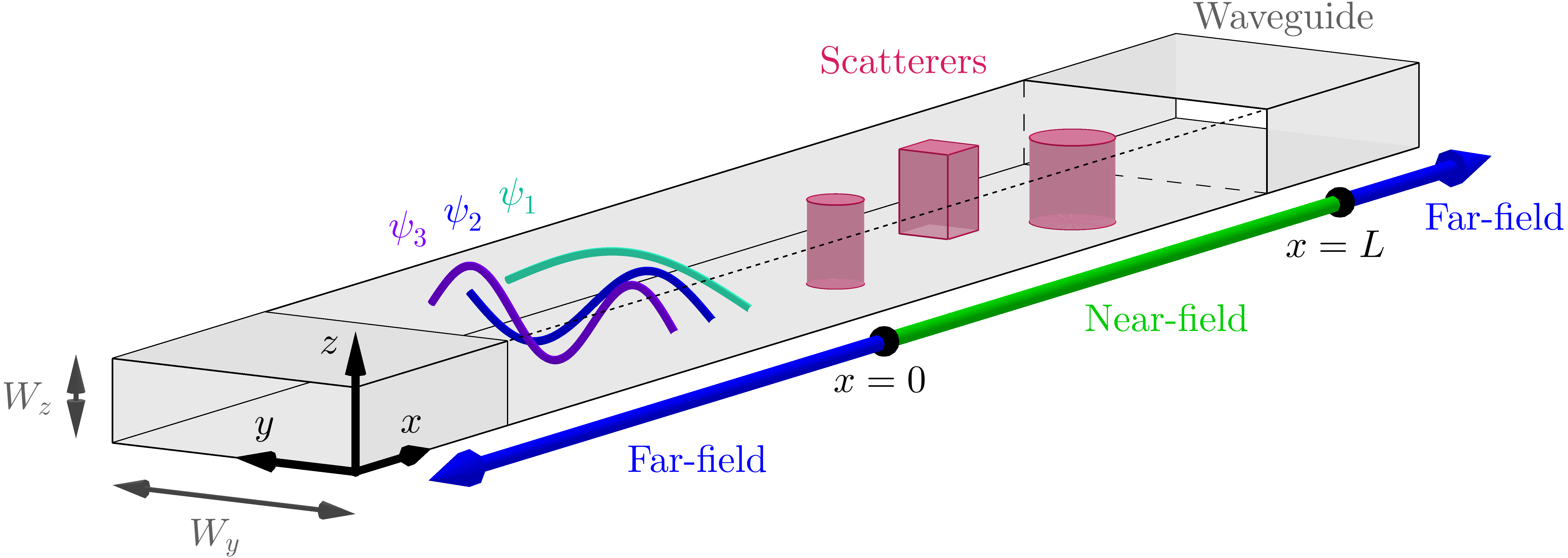}
    \caption{Electromagnetic waveguide (grey) with rectangular cross-section $W_y\times W_z$, extending along the $x$-axis. A section of the front side wall and the top plate are not shown to reveal a view of the interior. Scatterers (red) with different shapes and refractive indices constitute a scattering landscape inside the waveguide. The near-field (green) covers the spatial area where the scatterers are located. Its borders are defined as $x=0$ and $x=L$. The region outside the near-field is called the far-field (blue). The transverse profiles of the first three electromagnetic waveguide modes in $y$-direction, $\psi_1$, $\psi_2$ and $\psi_3$, are indicated in turquoise, blue and purple, respectively.}
    \label{fig:waveguide}
\end{figure*}

We start out by reviewing how the scattering of light can be formalized in classical optics. For such scattering processes, we distinguish the far-field from the near-field, see also Fig.\ \ref{fig:waveguide}. Objects that govern the non-trivial propagation of the wave and determine the scattering process are typically located in the near-field. The far-field, on the other hand, is characterized by free space propagation. The far-field is chosen sufficiently far away from the scattering region such that all evanescent waves have vanished. We assume here that light sources and detectors are placed in the far-field only.

If the far-field carries $N$ propagating modes (throughout this paper, we use the words ``channel'' and ``mode'' interchangeably), a wavefront that is injected into the system is described by $N$ amplitudes $\alpha_m^\mathrm{in}\in\mathbb{C}$, one for each mode $m\in\left\{1,\ldots,N\right\}$, or---more conveniently---by the vector $\bm{\alpha}^\mathrm{in}\in\mathbb{C}^N$. Likewise, the light that exits the system is described by the vector $\bm{\alpha}^\mathrm{out}\in\mathbb{C}^N$. Here, the amplitudes $\alpha_m$ are defined w.r.t.\ a flux-normalized basis \cite{RotterGigan2017}. The energy flux of the state $\bm{\alpha}$ is given by $\left\Vert\bm{\alpha}\right\Vert^2$. In an experiment, a detector typically measures the number of photons hitting the camera surface, which corresponds to the integrated energy flux \cite{RotterGigan2017}. For this reason, we will call $\left\Vert\bm{\alpha}\right\Vert^2$ the intensity of the state $\bm{\alpha}$.

The system is said to be linearly scattering if the media that the objects are made of have linear constitutive equations. Many ``ordinary'' optical elements are linear, e.g., mirrors, lenses, prisms, gratings, cavities and optical fibers. A linear scattering system is fully characterized by the frequency-dependent, so-called (classical) scattering matrix $\mathbf{S}\left(\omega\right)\in\mathbb{C}^{N\times N}$, which maps incoming monochromatic states of light $\bm{\alpha}^{\mathrm{in}}\left(\omega\right)$ to the corresponding output states $\bm{\alpha}^{\mathrm{out}}\left(\omega\right)$ \cite{RotterGigan2017}:
\begin{equation}
    \bm{\alpha}^{\mathrm{out}}\left(\omega\right)=\mathbf{S}\left(\omega\right)\bm{\alpha}^{\mathrm{in}}\left(\omega\right)\,.\label{eq:Smat_def}
\end{equation}
In absence of loss and gain, the scattering matrix is unitary and the output intensity equals the input intensity:
\begin{equation}
    \mathbf{S}^\dagger\left(\omega\right)\mathbf{S}\left(\omega\right)=\mathbf{1}\implies\left\Vert\bm{\alpha}^{\mathrm{out}}\left(\omega\right)\right\Vert^2=\left\Vert\bm{\alpha}^{\mathrm{in}}\left(\omega\right)\right\Vert^2\,.\label{eq:Smat_unitary}
\end{equation}

Experimentally, the optical transmission and reflection matrices, which are sub-parts of the scattering matrix, have been measured already \cite{Popoff2010_MeasureT,RahimiKeshari2013,Yu2013,Choi2013,Dremeau2015,Cao2023}.

Standard sources of classical light are lasers. Forming a given spatially shaped input state $\bm{\alpha}^{\mathrm{in}}\left(\omega\right)$ is called ``wavefront shaping'' \cite{Mosk2012,RotterGigan2017}. This is achieved by employing tools like spatial light modulators, digital micromirror devices or deformable mirrors in combination with lenses \cite{Vellekoop2007,Morizur2010,Mosk2012,Horstmeyer2015,Ayoub2021,Gigan2022,Cao2022}. Conventional detectors are charge-coupled device cameras. Homodyne detection provides a scheme to measure both the amplitude and the phase of an optical light field \cite{Lvovsky2009}.

The ideas and concepts that are discussed in this tutorial are kept very general and therefore apply to a wide range of optical systems. For illustration purposes, we select a concrete physical system, consisting of an infinite metallic wave\-guide along the $x$-axis with cross-section $W_y\times W_z$, see Fig.\ \ref{fig:waveguide}. The interior is filled with an isotropic, time- and $z$-independent medium described by the scalar electric susceptibility $\chi_\mathrm{e}\left(\omega;x,y\right)$. We assume a vanishing magnetic susceptibility $\chi_\mathrm{m}=0$, such that the refractive index landscape is given by $n\left(\omega;x,y\right)=\sqrt{1+\chi_\mathrm{e}\left(\omega;x,y\right)}$. Apart from this, there are no free charges or currents. We consider only monochromatic waves with frequency $\omega$ and wavenumber $k=\omega/c$. Furthermore, we demand that the waveguide is so narrow in the $z$-direction that only $\mathrm{TE}_{m_y,m_z}$ modes with $m_z=0$ can propagate. This is the case whenever $W_z<\pi/k$. In the following, we write $W\equiv W_y$ and $m\equiv m_y$. The $\mathrm{TE}_{m,0}$ modes are independent of the $z$ coordinate and they are polarized in the $z$-direction,
\begin{equation}
    \mathbf{E}\left(\mathbf{r},t\right)=\psi\left(\omega;x,y\right)\mathrm{e}^{-\mathrm{i}\omega t}\mathbf{e}_z\,.
\end{equation}
Inserting this field into Maxwell's equations yields the scalar two-dimensional Helmholtz equation
\begin{equation}
    \left(\partial_x^2+\partial_y^2+k^2n^2\left(\omega;x,y\right)\right)\psi\left(\omega;x,y\right)=0\label{eq:Helmholtz_2D}
\end{equation}
with the boundary conditions
\begin{align}
    \psi\left(\omega;x,0\right)&=0\,,\label{eq:Helmholtz_BC1}\\
    \psi\left(\omega;x,W\right)&=0\,.\label{eq:Helmholtz_BC2}
\end{align}

In the far-field, where $n\left(\omega;x,y\right)=1$, the waveguide modes are given by
\begin{equation}
    \psi_m^\pm\left(\omega;x,y\right)=\sqrt{\frac{2}{W}}\sin\left(\frac{m\pi y}{W}\right)\frac{\mathrm{e}^{\pm\mathrm{i}k_m^xx}}{\sqrt{k^x_m}}\,,\label{eq:wg_mode_full_xy}
\end{equation}
where the sign $\pm$ indicates the direction of travel ($+/-$ for propagation in positive/negative $x$ direction) and
\begin{equation}
    k_m^x=\sqrt{k^2-\frac{m^2\pi^2}{W^2}}
\end{equation}
is the wavenumber in the direction of propagation. The mode $m$ is called ``open'' or ``propagating'' when $k^x_m$ is real, i.e., $m<kW/\pi$. The mode $m$ is called ``evanescent'' when $k^x_m$ is imaginary, i.e., $m>kW/\pi$. Such modes decay exponentially fast and are not able to propagate into the far-field. For a given frequency $\omega=ck$ and waveguide width $W$, there are $N':=\lfloor\omega W/\pi c\rfloor$ open modes for each direction of propagation.

In the near-field, the scattering medium with $n\left(\omega;x,y\right)\neq 1$ should lie within $x\in\left[0,L\right]$. The two lines $x=0$ and $x=L$ mark the transition from the near-field to the far-field and serve as references for the scattering matrix (i.e., the numerical entries of $\mathbf{S}$ will depend on the choice of $L$). In the far-field, a general solution of Eqs.\ (\ref{eq:Helmholtz_2D}), (\ref{eq:Helmholtz_BC1}) and (\ref{eq:Helmholtz_BC2}) can be decomposed according to ($l$/$r$ refers to the left/right lead, i.e., $x\leq0$/$x\geq L$)
\begin{align}
    \psi\left(\omega;x\leq 0,y\right)&=\sum_{m=1}^{N'}\alpha_{l,m}^+\left(\omega\right)\psi_m^+\left(\omega;x,y\right)+\sum_{m=1}^{N'}\alpha_{l,m}^-\left(\omega\right)\psi_m^-\left(\omega;x,y\right)\,,\hspace{10mm}\label{eq:psi_mode_decomp_left}\\
    \psi\left(\omega;x\geq L,y\right)&=\sum_{m=1}^{N'}\alpha_{r,m}^+\left(\omega\right)\psi_m^+\left(\omega;x-L,y\right)+\sum_{m=1}^{N'}\alpha_{r,m}^-\left(\omega\right)\psi_m^-\left(\omega;x-L,y\right)\,.\label{eq:psi_mode_decomp_right}
\end{align}
The input wave is composed of right-travelling modes in the left lead and left-travelling modes in the right lead. For the output wave, the directions of travel are reversed. The input and output amplitudes are collected in the vectors
\begin{equation}
    \bm{\alpha}^{\mathrm{in}}\left(\omega\right)=
    \begin{pmatrix}
        \bm{\alpha}_l^+\left(\omega\right)\\
        \bm{\alpha}_r^-\left(\omega\right)
    \end{pmatrix}\,,\quad\bm{\alpha}^{\mathrm{out}}\left(\omega\right)=
    \begin{pmatrix}
        \bm{\alpha}_l^-\left(\omega\right)\\
        \bm{\alpha}_r^+\left(\omega\right)
    \end{pmatrix}\,,
\end{equation}
respectively. In total, there are $N=2N'$ input and output modes. The scattering matrix $\mathbf{S}\left(\omega\right)$ is defined through Eq.\ (\ref{eq:Smat_def}). The factor $1/\sqrt{k_m^x}$ in Eq.\ (\ref{eq:wg_mode_full_xy}) is necessary in order for the modes $\psi_m^\pm$ to be normalized w.r.t.\ the longitudinal flux \cite{RotterGigan2017}. The scattering matrix $\mathbf{S}\left(\omega\right)$ is unitary only with this normalization.

\begin{figure}[t!]
    \centering
    \subcaptionbox{Geometrical configuration\label{fig:system_config}}{
        \makebox[0.9\linewidth]{\includegraphics[scale=0.9]{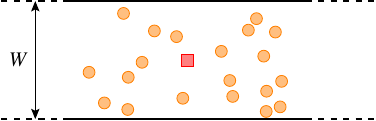}}
    }
    \vspace{3mm}
    \\
    \subcaptionbox{Scattering matrix\label{fig:system_Smat}}{
        \makebox[0.9\linewidth]{\includegraphics[scale=0.9]{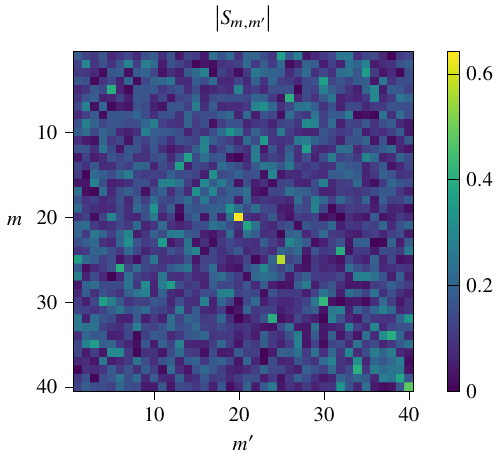}}
    }
    \caption{Physical setup of the waveguide system. \textbf{(a)} A metallic target (red square) is positioned inside a wave\-guide of width $W$. The target is surrounded by randomly placed circular scatterers (orange) with refractive index $n=1.44$. \textbf{(b)} Absolute values of the entries of the numerically calculated scattering matrix for the wavenumber $k=20.5\pi/W$, where $N=40$ waveguide modes are open.}
    \label{fig:system}
\end{figure}

A complex scattering system, as the one we are interested in here, is realized, e.g., by placing several scattering elements inside the waveguide, see Figs.\ \ref{fig:waveguide} and \ref{fig:system_config}. Let us consider a target scatterer in the shape of a square with side length $W/10$ which is positioned at the center of the waveguide. The target is metallic, which means that we impose homogeneous Dirichlet boundary conditions along its border. A complex scattering environment is provided by surrounding the target with 20 randomly placed circular scatterers with radius $W/20$. The refractive index of those scatterers is chosen as $1.44$. We numerically calculate the scattering matrix of this system for the wavenumber $k=20.5\pi/W$ such that $N'=20$ modes are open for each direction of propagation, i.e., there are $N=40$ open modes in total. Figure \ref{fig:system_Smat} shows the absolute values of the entries of the scattering matrix, their seemingly random distribution being a hallmark for complex scattering.

\subsection{Fundamentals of quantum optics\label{subsec:Fund_quant_optics}}

The classical magnetic vector potential $\mathbf{A}$ is usually decomposed in a suitable basis of modes. For the monochromatic Fourier component at frequency $\omega>0$, this decomposition reads
\begin{equation}
    \mathbf{A}\left(\mathbf{r},\omega\right)=\sum_{m=1}^Na_m\left(\omega\right)\mathbf{A}_m\left(\mathbf{r},\omega\right)\,.
\end{equation}
In the following, we consider only a single frequency component and thus omit to write $\omega$. The so-called ``second quantization'' or ``canonical quantization'' of electromagnetic radiation consists in replacing the coefficients $a_m$ by the annihilation operators $\hat{a}_m$ of the respective modes:
\begin{equation}
    \hat{\mathbf{A}}\left(\mathbf{r}\right)=\sum_{m=1}^N\hat{a}_m\mathbf{A}_m\left(\mathbf{r}\right)\,.
\end{equation}
The Hermitian conjugates $\hat{a}_m^\dagger$ of the annihilation operators are called ``creation operators''. The operators fulfil the bosonic commutation relations
\begin{equation}
    \left[\hat{a}_m,\hat{a}_{m'}^\dagger\right]=\delta_{m,m'}\,,\quad \left[\hat{a}_m,\hat{a}_{m'}\right]=0\,,\quad \left[\hat{a}_m^\dagger,\hat{a}_{m'}^\dagger\right]=0\,.\label{eq:bose_commutators}
\end{equation}
Physically speaking, the annihilation and creation operators destroy and create single quanta of light, i.e.\ photons, respectively. We denote the column vector consisting of the annihilation operators by
\begin{equation}
    \left[\hat{a}\right]:=\left(\hat{a}_1,\hat{a}_2,\ldots,\hat{a}_N\right)^\top
\end{equation}
and likewise $\left[\hat{a}^\dagger\right]$ for the creation operators. The observables
\begin{align}
    \hat{n}_m&:=\hat{a}_m^\dagger\hat{a}_m\,,\label{eq:n_m_def}\\
    \hat{n}&:=\sum_{m=1}^N\hat{n}_m\label{eq:n_total_def}
\end{align}
measure the number of photons in mode $m$ and the total number of photons, respectively. The Hamiltonian of the electromagnetic field turns out to be
\begin{equation}
    \hat{H}_\mathrm{EM}=\int\mathrm{d}\omega\sum_{m=1}^N\hbar\omega\left(\hat{n}_m+\frac{1}{2}\right)\,.\label{eq:Hamiltonian_EM}
\end{equation}
This means that the quantized electromagnetic field can be pictured as a collection of independent harmonic oscillators, one for each mode at each frequency. Correspondingly, multi-mode light can take the same quantum states as a multi-dimensional harmonic oscillator. Similar to the position and momentum of a harmonic oscillator, the so-called ``quadratures'' of mode $m$ are introduced as
\begin{align}
    \hat{q}_m&:=\frac{1}{\sqrt{2}}\left(\hat{a}_m+\hat{a}_m^\dagger\right)\,,\\
    \hat{p}_m&:=\frac{-\mathrm{i}}{\sqrt{2}}\left(\hat{a}_m-\hat{a}_m^\dagger\right)\,.
\end{align}
A rotation in the phase space spanned by $\hat{q}_m$ and $\hat{p}_m$ yields the ``rotated'' quadratures (also often just called quadratures)
\begin{align}
    \hat{q}_m\left(\varphi\right)&:=\cos\left(\varphi\right)\hat{q}_m+\sin\left(\varphi\right)\hat{p}_m\nonumber\\
    &=\frac{1}{\sqrt{2}}\left(\mathrm{e}^{-\mathrm{i}\varphi}\hat{a}_m+\mathrm{e}^{\mathrm{i}\varphi}\hat{a}_m^\dagger\right)\,,\label{eq:rotated_q_def}\\
    \hat{p}_m\left(\varphi\right)&:=-\sin\left(\varphi\right)\hat{q}_m+\cos\left(\varphi\right)\hat{p}_m\nonumber\\
    &=\frac{-\mathrm{i}}{\sqrt{2}}\left(\mathrm{e}^{-\mathrm{i}\varphi}\hat{a}_m-\mathrm{e}^{\mathrm{i}\varphi}\hat{a}_m^\dagger\right)\,.\label{eq:rotated_p_def}
\end{align}

Below, we discuss a physically relevant selection of quantum states of light with $N$ modes. To make it clear that the representation is rooted in the ``standard'' modes (e.g., the waveguide modes (\ref{eq:wg_mode_full_xy})), we use the superscript $\mathcal{M}$ in the notation. Later on, we will use other representations as well.

Coherent states $\ket{\bm{\alpha}}^\mathcal{M}$ are parameterized by a vector of coherent amplitudes $\bm{\alpha}\in\mathbb{C}^N$. Introducing the following unitary displacement operator,
\begin{equation}
    \hat{D}_a\left(\bm{\alpha}\right):=\exp\left(\bm{\alpha}^\top\left[\hat{a}^\dagger\right]-\bm{\alpha}^\dagger\left[\hat{a}\right]\right)\,,\label{eq:Displacement_Op}
\end{equation}
the coherent state $\ket{\bm{\alpha}}^\mathcal{M}$ is created from the vacuum state $\ket{\mathbf{0}}$ via displacement, see also Fig.\ \ref{fig:Coh_state}:
\begin{figure*}[t]
    \subcaptionbox{Coherent state\label{fig:Coh_state}}{
        \includegraphics[scale=0.9]{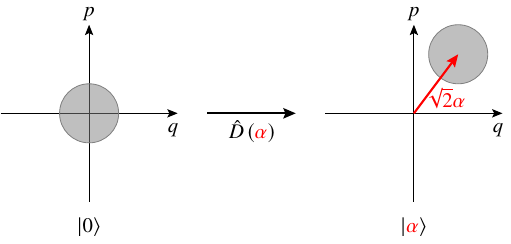}
    }
    \vspace{3mm}
    \\
    \subcaptionbox{Squeezed state\label{fig:Squ_state}}{
        \includegraphics[scale=0.9]{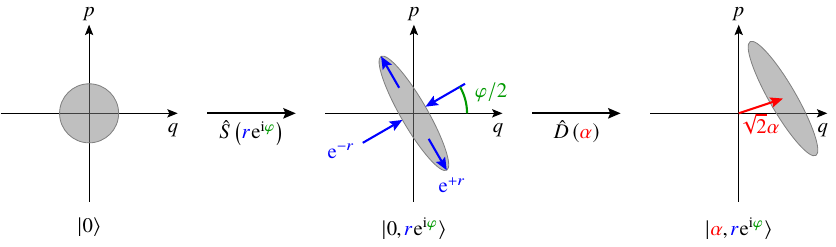}
    }
    \caption{Phase space representation of single-mode Gaussian states of light. The grey circles and ellipses represent contour sections of the Wigner functions. \textbf{(a)} A coherent state $\ket{\alpha}$ is created from the vacuum state $\ket{0}$ by applying the displacement operator $\hat{D}\left(\alpha\right)$. \textbf{(b)} A squeezed state $\ket{\alpha,z}$ with $z=r\mathrm{e}^{\mathrm{i}\varphi}$ is created from the vacuum state $\ket{0}$ by first applying the squeezing operator $\hat{S}\left(z\right)$ and then the displacement operator $\hat{D}\left(\alpha\right)$. Here, the squeezing parameters are chosen as $r=0.8$ and $\varphi=\ang{60}=\pi/3$.}
    \label{fig:Wigner_func}
\end{figure*}
\begin{equation}
    \ket{\bm{\alpha}}^\mathcal{M}:=\hat{D}_a\left(\bm{\alpha}\right)\ket{\mathbf{0}}\,.
\end{equation}
Coherent states are eigenstates of the annihilation operators,
\begin{equation}
    \hat{a}_m\ket{\bm{\alpha}}^\mathcal{M}=\alpha_m\ket{\bm{\alpha}}^\mathcal{M}\,,
\end{equation}
and they have the mean photon numbers
\begin{align}
    \tensor[^{\mathcal{M}}]{\braket{\bm{\alpha}|\hat{n}_m|\bm{\alpha}}}{^{\mathcal{M}}}&=\left|\alpha_m\right|^2\,,\\
    \tensor[^{\mathcal{M}}]{\braket{\bm{\alpha}|\hat{n}|\bm{\alpha}}}{^{\mathcal{M}}}&=\left\Vert\bm{\alpha}\right\Vert^2\,.
\end{align}
They exhibit equal minimum uncertainty (i.e., variance $\mathbb{V}$) in all quadratures,
\begin{align}
    &\forall m\in\left\{0,\ldots,N\right\}\forall\varphi\in\left[0,2\pi\right):\mathbb{V}_{\ket{\bm{\alpha}}^\mathcal{M}}\left[\hat{q}_m\left(\varphi\right)\right]\equiv\nonumber\\
    &\equiv\tensor[^{\mathcal{M}}]{\braket{\bm{\alpha}|\hat{q}_m^2\left(\varphi\right)|\bm{\alpha}}}{^{\mathcal{M}}}-\left(\tensor[^{\mathcal{M}}]{\braket{\bm{\alpha}|\hat{q}_m\left(\varphi\right)|\bm{\alpha}}}{^{\mathcal{M}}}\right)^2=\frac{1}{2}\,,
\end{align}
which makes them the ``most classical'' states of quantum light. Coherent states $\ket{\bm{\alpha}}^\mathcal{M}$ can thus be identified with the classical light states $\bm{\alpha}$ discussed in Subsection \ref{subsec:Scat_cl_optics}.

Squeezed states $\ket{\bm{\alpha},\mathbf{Z}}^\mathcal{M}$ are parameterized by a vector of coherent amplitudes $\bm{\alpha}\in\mathbb{C}^N$ and a symmetric squeezing matrix $\mathbf{Z}\in\mathbb{C}^{N\times N}$ \cite{MaRhodes1990}. Introducing the unitary squeezing operator (note the different sign convention in Ref.\ \cite{MaRhodes1990})
\begin{equation}
    \hat{S}_a\left(\mathbf{Z}\right):=\exp\left(\frac{1}{2}\left(\left[\hat{a}\right]^\top\mathbf{Z}^\ast\left[\hat{a}\right]-\left[\hat{a}^\dagger\right]^\top\mathbf{Z}\left[\hat{a}^\dagger\right]\right)\right)\,,\label{eq:Squeezing_Op}
\end{equation}
the squeezed state $\ket{\bm{\alpha},\mathbf{Z}}^\mathcal{M}$ is obtained from the vacuum state by squeezing it by $\mathbf{Z}$ first and then displacing it by $\bm{\alpha}$, see also Fig.\ \ref{fig:Squ_state}:
\begin{equation}
    \ket{\bm{\alpha},\mathbf{Z}}^\mathcal{M}:=\hat{D}_a\left(\bm{\alpha}\right)\hat{S}_a\left(\mathbf{Z}\right)\ket{\mathbf{0}}\,.
\end{equation}
The Wigner function of a squeezed state is a Gaussian \cite{Leonhardt2010}. For this reason, squeezed states (including coherent states) are also often called ``Gaussian states''.

We now discuss a specific decomposition of the squeezing matrix $\mathbf{Z}$ which will be useful in later calculations. This decomposition is the counterpart to the polar representation of a scalar complex number, $z=r\mathrm{e}^{\mathrm{i}\varphi}$, and is thus called ``polar decomposition'': Any finite-dimensional square matrix can be decomposed into a product of a Hermitian and a unitary matrix \cite{MaRhodes1990}:
\begin{equation}
    \mathbf{Z}=\mathbf{R}\mathrm{e}^{\mathrm{i}\bm{\Phi}}\,.\label{eq:Polar_decomp_Z}
\end{equation}
Both matrices $\mathbf{R}$ and $\bm{\Phi}$ are Hermitian with the properties $\mathbf{0}\preceq\mathbf{R}$, $\mathbf{0}\prec\mathbf{R}\iff\det\left(\mathbf{Z}\right)\neq0$, and $\mathbf{0}\preceq\bm{\Phi}\prec2\pi\mathbf{1}$. For Hermitian matrices $\mathbf{A},\mathbf{B}$ we denote $\mathbf{A}\prec\mathbf{B}$ if $\mathbf{B}-\mathbf{A}$ is positive definite and $\mathbf{A}\preceq\mathbf{B}$ if $\mathbf{B}-\mathbf{A}$ is positive semidefinite. $\mathbf{R}$ is always unique, but $\bm{\Phi}$ is unique iff $\det\left(\mathbf{Z}\right)\neq0$. The polar decomposition can be obtained from the singular value decomposition $\mathbf{Z}=\mathbf{U}\bm{\Sigma}\mathbf{V}^\dagger$ as $\mathbf{R}=\mathbf{U}\bm{\Sigma}\mathbf{U}^\dagger$ and $\mathrm{e}^{\mathrm{i}\bm{\Phi}}=\mathbf{U}\mathbf{V}^\dagger$.

In Appendix \ref{sec:App_QFI_Gaussian} we show that the mean photon numbers of a squeezed state are
\begin{align}
    \tensor[^{\mathcal{M}}]{\braket{\bm{\alpha},\mathbf{Z}|\hat{n}_m|\bm{\alpha},\mathbf{Z}}}{^{\mathcal{M}}}&=\left|\alpha_m\right|^2+\left(\sinh^2\left(\mathbf{R}\right)\right)_{m,m}\,,\label{eq:mean_nm_squeezed}\\
    \tensor[^{\mathcal{M}}]{\braket{\bm{\alpha},\mathbf{Z}|\hat{n}|\bm{\alpha},\mathbf{Z}}}{^{\mathcal{M}}}&=\left\Vert\bm{\alpha}\right\Vert^2+\tr\left(\sinh^2\left(\mathbf{R}\right)\right)\,,\label{eq:mean_n_squeezed}
\end{align}
where $\sinh^2$ is to be applied as a proper matrix function.

If the squeezing matrix is diagonal, $\mathbf{Z}=\diag\left(\zeta_1,\ldots,\zeta_N\right)$ with $\zeta_m=r_m\mathrm{e}^{\mathrm{i}\varphi_m}$, the squeezed state factorizes into a product of single-mode squeezed states. In phase space, the physical interpretation of the parameters $r_m$ and $\varphi_m/2$ is squeezing strength and squeezing angle, respectively, see also Fig.\ \ref{fig:Squ_state}. The squeezing strength determines the uncertainties in the rotated quadratures (a proof is omitted here):
\begin{align}
    \mathbb{V}_{\ket{\bm{\alpha},\mathbf{Z}}^\mathcal{M}}\left[\hat{q}_m\left(\varphi_m/2\right)\right]&=\frac{\mathrm{e}^{-2r_m}}{2}\,,\label{eq:Var_sqz_q}\\
    \mathbb{V}_{\ket{\bm{\alpha},\mathbf{Z}}^\mathcal{M}}\left[\hat{p}_m\left(\varphi_m/2\right)\right]&=\frac{\mathrm{e}^{2r_m}}{2}\,.\label{eq:Var_sqz_p}
\end{align}
Squeezing strengths ($r$) are often stated in units of decibel, the conversion being $\frac{20}{\ln\left(10\right)}r\,\si{\decibel}\approx 8.686r\,\si{\decibel}$. The highest squeezing strength currently accomplished experimentally is $\SI{15}{\decibel}$ or $r\approx 1.73$ \cite{Vahlbruch2016}.

\subsection{Scattering in quantum optics\label{subsec:Scat_qu_optics}}

Analogously to Eq.\ (\ref{eq:Smat_def}), which describes unitary classical scattering, we declare a unitary quantum scattering process to be defined by a unitary operator $\hat{U}$ mapping pure input states to pure output states:
\begin{equation}
    \ket{\psi^\mathrm{out}}=\hat{U}\ket{\psi^\mathrm{in}}\,.\label{eq:U_def}
\end{equation}
This operator $\hat{U}$ is determined by the input-output relation of the mode operators $\hat{a}_m$ and $\hat{a}_m^\dagger$. We assume a linear input-output relation, allowing for annihilation operators to be transformed into annihilation operators only. This covers all passive linear elements as discussed in Subsection \ref{subsec:Scat_cl_optics} \cite{Leonhardt2003,LeonhardtNeumaier2003}. For optomechanical and micromechanical systems which do not satisfy this condition, our framework is restricted to the linear regime (as realized, e.g., for sufficiently low intensities). In formal terms, such a transformation reads:
\begin{equation}
    \left[\hat{U}^\dagger\hat{a}\hat{U}\right]=\mathbf{A}\left[\hat{a}\right]\,.\label{eq:Bogoliubov_trafo}
\end{equation}
As Maxwell's equations hold both for classical and quantum fields, the quantum amplitude must transform in exactly the same way as the classical ones. Hence, the matrix $\mathbf{A}$ must be the classical scattering matrix $\mathbf{S}$. This is confirmed by Eq.\ (\ref{eq:U_coherent}). The unitarity (\ref{eq:Smat_unitary}) of $\mathbf{S}$ ensures that the transformed operators still fulfil the commutation relations (\ref{eq:bose_commutators}). In terms of $\mathbf{S}$, the unitary operator $\hat{U}$ is given as \cite{MaRhodes1990,Leonhardt2003,LeonhardtNeumaier2003}
\begin{equation}
    \hat{U}=\sqrt{\det\left(\mathbf{S}\right)}\exp\left(\left[\hat{a}^\dagger\right]^\top\ln\left(\mathbf{S}\right)\left[\hat{a}\right]\right)\,.\label{eq:U_S}
\end{equation}
This relation yields the main insight of this subsection, namely that the scattering behavior of multi-mode quantum light is fully determined by the classical scattering matrix alone. In general, the task of determining a quantum unitary gate by probing it with different (often coherent) states is called ``quantum process tomography'' \cite{Lobino2008,RahimiKeshari2011,RahimiKeshari2013,Jacob2018}.

A first important observation is that such a quantum process does not change the total photon number, i.e., $\hat{n}\hat{U}=\hat{U}\hat{n}$. This can be shown by using Eqs.\ (\ref{eq:Smat_unitary}), (\ref{eq:n_m_def}), (\ref{eq:n_total_def}) and (\ref{eq:Bogoliubov_trafo}):
\begin{align}
    \hat{U}^\dagger\hat{n}\hat{U}&=\hat{U}^\dagger\left[\hat{a}^\dagger\right]^\top\left[\hat{a}\right]\hat{U}=\left[\hat{U}^\dagger\hat{a}^\dagger\hat{U}\right]^\top\left[\hat{U}^\dagger\hat{a}\hat{U}\right]=\left(\mathbf{S}^\ast\left[\hat{a}^\dagger\right]\right)^\top\mathbf{S}\left[\hat{a}\right]=\left[\hat{a}^\dagger\right]^\top\mathbf{S}^\dagger\mathbf{S}\left[\hat{a}\right]\nonumber\\
    &=\left[\hat{a}^\dagger\right]^\top\left[\hat{a}\right]=\hat{n}\,.
\end{align}

According to Ref.\ \cite{MaRhodes1990}, the scattering behavior of a Gaussian state is given by
\begin{equation}
    \hat{U}\ket{\bm{\alpha},\mathbf{Z}}^\mathcal{M}=\sqrt{\det\left(\mathbf{S}\right)}\ket{\mathbf{S}\bm{\alpha},\mathbf{S}\mathbf{Z}\mathbf{S}^\top}^\mathcal{M}\,.\label{eq:U_squeezed}
\end{equation}
Coherent states are transformed according to the classical scattering matrix:
\begin{equation}
    \hat{U}\ket{\bm{\alpha}}^\mathcal{M}=\sqrt{\det\left(\mathbf{S}\right)}\ket{\mathbf{S}\bm{\alpha}}^\mathcal{M}\,.\label{eq:U_coherent}
\end{equation}

\section{Wigner-Smith Formalism\label{sec:WS_Formalism}}

\subsection{The classical Wigner-Smith matrix\label{subsec:ClWSMat}}

As was first shown by Eisenbud, Wigner and Smith \cite{Eisenbud1948,Wigner1955,Smith1960}, the scattering matrix provides access to the time spent by waves in the scattering process through the Wigner-Smith time-delay matrix, 
\begin{equation}
    \mathbf{Q}_\omega:=-\mathrm{i}\mathbf{S}^\dagger\left(\omega\right)\partial_\omega\mathbf{S}\left(\omega\right)\,,\label{eq:WS_def}
\end{equation}
involving a frequency derivative of the scattering matrix. The so-called ``proper delay times'' are defined as the eigenvalues of this Hermitian matrix \cite{Winful2003,Kottos2005} and the corresponding eigenstates, also known as ``principal modes'' \cite{Shanhui2005}, are the input vectors (in the mode basis) for the scattering states associated with these well-defined delay times. This concept can be generalized to involve, instead of the frequency derivative, a derivative with respect to any other parameter $\theta$ that the scattering matrix depends on. Instead of the time-delay---as the conjugate quantity to the frequency---such a generalized Wigner-Smith (GWS) matrix, defined as follows
\begin{equation}
    \mathbf{Q}_\theta:=-\mathrm{i}\mathbf{S}^\dagger\left(\theta\right)\partial_\theta\mathbf{S}\left(\theta\right)\,,\label{eq:GWS_def}
\end{equation}
then provides access to the physical observable associated with the quantity conjugate to $\theta$ \cite{Ambichl2017_Focusing,Horodynski2020}. Let $\theta$ be, e.g., the position or the rotation angle of a target. In this case, the expectation value $\bm{\alpha}^\dagger\mathbf{Q}_\theta\bm{\alpha}$ is proportional to the mean force or torque, respectively, acting on this target in the direction of increasing $\theta$ by the input state $\bm{\alpha}$. We call these opto-mechanical forces and torques (and other transfers of quantities conjugate to some $\theta$) ``generalized forces''.

By definition, the GWS matrix is Hermitian if the scattering matrix is unitary:
\begin{align}
    &\partial_\theta\left(\mathbf{S}^\dagger\mathbf{S}\right)=\left(\partial_\theta\mathbf{S}\right)^\dagger\mathbf{S}+\mathbf{S}^\dagger\partial_\theta\mathbf{S}=\mathbf{0}\nonumber\\
    &\implies\mathbf{Q}_\theta^\dagger=\mathrm{i}\left(\partial_\theta\mathbf{S}\right)^\dagger\mathbf{S}=-\mathrm{i}\mathbf{S}^\dagger\partial_\theta\mathbf{S}=\mathbf{Q}_\theta\,.
\end{align}
This implies that the eigenvalues $\lambda_i$ of $\mathbf{Q}_\theta$ are real-valued and the corresponding eigenvectors $\mathbf{w}_i$ form an orthonormal basis of $\mathbb{C}^N$:
\begin{equation}
    \mathbf{Q}_\theta\mathbf{w}_i=\lambda_i\mathbf{w}_i\iff\mathbf{Q}_\theta=\mathbf{W}\bm{\Lambda}\mathbf{W}^\dagger\,.\label{eq:Q_eigdecomp}
\end{equation}
The eigenvectors $\mathbf{w}_i$ are the input states that deliver a certain generalized force conjugate to $\theta$ that is proportional to the corresponding eigenvalue $\lambda_i$. Consequently, the eigenvector of the GWS matrix with the largest eigenvalue provides the incoming wavefront that couples to the parameter $\theta$ most strongly and thus constitutes the optimal wave state for micromanipulating this target. The GWS matrix can, however, also be applied in a broader context such as for the optimal cooling of an ensemble of particles \cite{Huepfl2023}, for the identification of channels that are resilient to disorder \cite{Matthes2021}, or for the optimal retrieval of information on the system parameter $\theta$ in an arbitrarily complex scattering environment \cite{Bouchet2021}.

The purpose of the GWS matrix is to provide access to the relevant quantities for manipulating a target without knowledge of the target's near-field; only the scattering amplitudes in the far-field and their dependence on the relevant parameter $\theta$ are required. Since no direct access to the target scatterer is necessary in this way, this target may also be hidden behind or inside a complex medium like a disordered material. Note that, for accessing the $\theta$-dependence of the scattering matrix, a small, controlled variation of $\theta$ must occur in the system. Experimentally, there are different ways of how this can be achieved, such as by externally induced forces (using acoustic, magnetic or gravitational fields \cite{Judkewitz2013,Zhou2014,Ma2014,Ruan2017}) or by autonomous movement \cite{Bechinger2016,Frangipane2019,Huepfl2023}.

\begin{figure*}[t]
    \centering
    \subcaptionbox{Maximum force to the right\label{fig:mm_h_max}}{
        \includegraphics[scale=0.88]{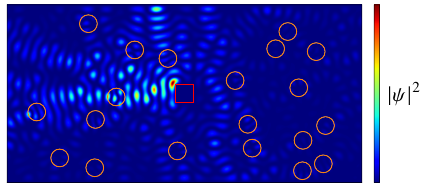}
    }
    \hfill
    \subcaptionbox{Maximum force to the left\label{fig:mm_h_min}}{
        \includegraphics[scale=0.88]{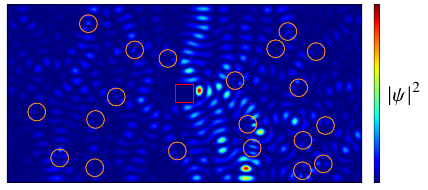}
    }
    \vspace{3mm}
    \\
    \subcaptionbox{Maximum force upwards\label{fig:mm_v_max}}{
        \includegraphics[scale=0.88]{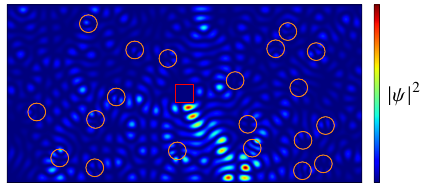}
    }
    \hfill
    \subcaptionbox{Maximum force downwards\label{fig:mm_v_min}}{
        \includegraphics[scale=0.88]{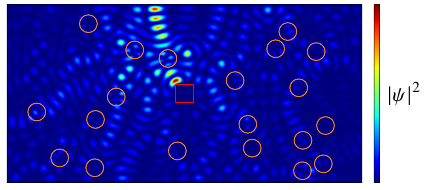}
    }
    \caption{Classical optical micromanipulation with the parameter of interest $\theta$ being the horizontal position $x$ or the vertical position $y$ of the target (red square in the center). The corresponding GWS matrices $\mathbf{Q}_\theta$ and their eigenvalues quantify the corresponding momentum transfers onto the target. The plots show the spatial intensity distributions of the emerging waves when specific eigenstates of the GWS matrices $\mathbf{Q}_\theta$ are injected into the system. In the regions of high intensity close to the target, the wave exerts a local force onto the target. \textbf{(a)} Eigenstate corresponding to the maximum eigenvalue of $\mathbf{Q}_x$. \textbf{(b)} Eigenstate corresponding to the minimum eigenvalue of $\mathbf{Q}_x$. \textbf{(c)} Eigenstate corresponding to the maximum eigenvalue of $\mathbf{Q}_y$. \textbf{(d)} Eigenstate corresponding to the minimum eigenvalue of $\mathbf{Q}_y$.}
    \label{fig:Optimal_micromanipulation}
\end{figure*}

To illustrate how the GWS matrix is employed for micromanipulation, we turn to the generic example introduced at the end of Subsection \ref{subsec:Scat_cl_optics}. For the parameter $\theta$ we choose two realizations that we consider separately: horizontal ($\theta=x$) and vertical ($\theta=y$) displacement of the target. In the numerical simulation, we use a finite difference approximation for the $\theta$-derivative. Figures \ref{fig:mm_h_max} and \ref{fig:mm_h_min} show the spatial intensity distribution of the wave that emerges when injecting eigenstates of $\mathbf{Q}_x$ into the system corresponding to the maximum and minimum eigenvalue, respectively. In the immediate vicinity of the target, regions of high intensity exert a force onto the target. It is apparent that the waves in Figs.\ \ref{fig:mm_h_max} and \ref{fig:mm_h_min} lead to a force pointing to the right and left, respectively. Likewise, Figs.\ \ref{fig:mm_v_max} and \ref{fig:mm_v_min} show the eigenstates of $\mathbf{Q}_y$ corresponding to the maximum and minimum eigenvalue, respectively.

\subsection{The quantum Wigner-Smith operator\label{subsec:QuWSOp}}

Let us now extend the scope of the GWS matrix from classical wave physics to quantum mechanics. To this end, we will reformulate the Wigner-Smith framework such as to apply it also to quantized electromagnetic fields. A central insight in this context is the result from Subsection \ref{subsec:Scat_qu_optics} that for any linear optical network, the classical description of scattering can be directly translated into an equivalent quantum evolution. In the same way as the scattering matrix $\mathbf{S}$ transfers the vector of classical input amplitudes to the output, see Eq.\ (\ref{eq:U_coherent}), the unitary operator $\hat{U}$ acts on the corresponding multi-mode input quantum state to yield the corresponding output quantum state, see Eq.\ (\ref{eq:U_def}). Correspondingly, we translate the GWS matrix $\mathbf{Q}_\theta$ from  Eq.\ (\ref{eq:GWS_def}) by replacing $\mathbf{S}$ with $\hat{U}$ to arrive at what we refer to as the quantum Wigner-Smith (QWS) operator:
\begin{equation}
    \hat{Q}_\theta:=-\mathrm{i}\hat{U}^\dagger\left(\theta\right)\partial_\theta\hat{U}\left(\theta\right)\,.\label{eq:QWS_def}
\end{equation}

In quantum metrology and Lie group theory, this operator is known as a generator and has already been studied in detail \cite{Das2014,Pang2017,Fiderer2019}. Here, $\hat{U}$ constitutes a linear Gaussian process, characterized by the classical scattering matrix $\mathbf{S}$. Note that the definition (\ref{eq:QWS_def}) of the QWS operator and its connection to the major quantities in this work, like the generalized force in Eq.\ (\ref{eq:K_int_Q}) and the quantum Fisher information in Eq.\ (\ref{eq:QFI_VarQ}), remain valid for processes that are not linear on the classical level, but still linear on the quantum level. Such processes, including so-called ``active linear elements'' like phase conjugation mirrors or parametric amplifiers, cannot be described by a classical scattering matrix, but by a unitary quantum operator \cite{Leonhardt2003,LeonhardtNeumaier2003}. Throughout the rest of this manuscript, we focus on processes that are linear on the classical level since in this case all the quantities are expressible in terms of the experimentally accessible scattering matrix.

To arrive at the main result of this work, we insert the transformation (\ref{eq:U_S}) into the definition (\ref{eq:QWS_def}) of the QWS operator (a detailed derivation is given in Appendix \ref{sec:App_proof_QWS_GWS}),
\begin{equation}
    \hat{Q}_\theta=\left[\hat{a}^\dagger\right]^\top\mathbf{Q}_\theta\left[\hat{a}\right]+\frac{1}{2}\tr\left(\mathbf{Q}_\theta\right)\,.\label{eq:QWS_GWS}
\end{equation}
This remarkably simple relation provides an operational procedure for translating the classical scattering amplitudes in the measurable GWS matrix $\mathbf{Q}_\theta$ to a corresponding quantum operator $\hat{Q}_\theta$. This is most directly seen in the first term on the right hand side of Eq.\ (\ref{eq:QWS_GWS}), which simply couples the elements of the classical GWS matrix to the corresponding quantum channels, similar to the Jordan-Schwinger map \cite{Jordan1935,Schwinger1952}. The QWS operator inherits some properties from the GWS matrix, like being Hermitian for unitary systems and its ability to express local phenomena in terms of the far-field scattering amplitudes. Moreover, as a result of the nonlinear relation in Eq.\ (\ref{eq:U_S}), the normal ordering in Eq.\ (\ref{eq:QWS_GWS}) and the non-commutativity of the mode operators, we do find an additional scalar trace term (the second term on the right-hand side of Eq.\ (\ref{eq:QWS_GWS})). As it turns out, this term is not connected to the force exerted by the injected field, but rather due to the forces of the quantum vacuum. This will be discussed in more detail in Section \ref{sec:VacForces}.

The QWS operator unites the spatial and the quantum degrees of freedom of scattered light fields such that we can describe and optimize them jointly to perform both micromanipulation and parameter estimation at the optimal level of efficiency, as will be detailed in the following Sections \ref{sec:QuMicroMan} and \ref{sec:QuMetrology}.

Before doing so, we introduce a new representation (in contrast to the modal representation $\mathcal{M}$) based on the eigen-decomposition (\ref{eq:Q_eigdecomp}) of the GWS matrix $\mathbf{Q}_\theta$. This new representation will be very useful for upcoming calculations. We indicate it with the symbol $\mathcal{Q}$. The channels that form the basis of the $\mathcal{Q}$ representation are given by the eigenvectors $\mathbf{w}_i$ of $\mathbf{Q}_\theta$. These eigenvectors are the columns of the unitary matrix $\mathbf{W}$. We denote the corresponding annihilation operators with $\hat{b}_i$ and their connection to the mode operators $\hat{a}_m$ is expressed as
\begin{equation}
    [\hat{b}]=\mathbf{W}^\dagger[\hat{a}]\,.\label{eq:Annih_b_def}
\end{equation}
Due to the unitarity of $\mathbf{W}$, the total photon number operator is independent of the representation (see Eq.\ (\ref{eq:n_total_def})),
\begin{equation}
    \hat{\nu}:=\sum_{i=1}^N\hat{b}_i^\dagger\hat{b}_i=[\hat{b}^\dagger]^\top[\hat{b}]=[\hat{a}^\dagger]^\top\mathbf{W}\mathbf{W}^\dagger[\hat{a}]=[\hat{a}^\dagger]^\top[\hat{a}]=\hat{n}\,.
\end{equation}
It is straightforward to show that the displacement operator (\ref{eq:Displacement_Op}) and the squeezing operator (\ref{eq:Squeezing_Op}) transform according to
\begin{align}
    \hat{D}_a\left(\bm{\alpha}\right)&=\hat{D}_b\left(\bm{\beta}:=\mathbf{W}^\dagger\bm{\alpha}\right)\,,\label{eq:DisplOp_trafo}\\
    \hat{S}_a\left(\mathbf{Z}\right)&=\hat{S}_b\left(\bm{\Xi}:=\mathbf{W}^\dagger\mathbf{Z}\mathbf{W}^\ast\right)\,.\label{eq:SqueezeOp_trafo}
\end{align}
We denote
\begin{equation}
    \ket{\bm{\beta},\bm{\Xi}}^{\mathcal{Q}}:=\hat{D}_b\left(\bm{\beta}\right)\hat{S}_b\left(\bm{\Xi}\right)\ket{\mathbf{0}}\,,\label{eq:SqCohState_def_Q}
\end{equation}
which is the same state as $\ket{\bm{\alpha},\mathbf{Z}}^{\mathcal{M}}$ but in a different representation. The polar decomposition (see Eq.\ (\ref{eq:Polar_decomp_Z})) $\bm{\Xi}=\mathbf{P}\mathrm{e}^{\mathrm{i}\bm{\Psi}}$ transforms according to $\mathbf{P}=\mathbf{W}^\dagger\mathbf{R}\mathbf{W}$ and $\mathrm{e}^{\mathrm{i}\bm{\Psi}}=\mathbf{W}^\dagger\mathrm{e}^{\mathrm{i}\bm{\Phi}}\mathbf{W}^\ast$.

\section{Quantum micromanipulation\label{sec:QuMicroMan}}

A first application of the QWS operator lies in micromanipulation, in which domain the quantum degrees of freedom of light have, e.g., been used already to improve cooling protocols \cite{Asjad2016,Clark2017,Asjad2019,Monsel2021}. Here, we are interested in finding a state of light that, when injected into the scattering system, optimally couples to the system property described by $\theta$, which can be any geometric or material parameter characterizing the scattering system as a whole or any part of it. For a given fixed mean photon number, which is proportional to the total energy of the incident light (apart from the zero point energy), we aim to identify input states that exert the highest generalized force in the direction of increasing $\theta$, when compared to all other possible input states with the same mean photon number.

In Appendix \ref{sec:App_Micromanip_PhysInterpret} we show that the multi-spectral quantum operator corresponding to the generalized force conjugate to $\theta$ is given by the following expression, which involves all frequency components in the entire electromagnetic spectrum:
\begin{equation}
    \hat{K}_\theta=\frac{1}{2\pi}\int_0^\infty\hat{Q}_\theta\,\mathrm{d}E\,.\label{eq:K_int_Q}
\end{equation}
Here, $E=\hbar\omega$ denotes the photonic energy corresponding to the frequency $\omega$, $\hbar$ being the reduced Planck constant. This relation shows that, for a broadband light field, the QWS operator is the spectral density of the generalized force.

Next, we insert Eq.\ (\ref{eq:QWS_GWS}) into Eq.\ (\ref{eq:K_int_Q}) and take the expectation value with respect to some input state of light $\ket{\Psi_0}$. It is to be understood that $\ket{\Psi_0}$ is composed of spectral components $\ket{\psi_E}$ from the whole energy spectrum with respective amplitudes $c(E)$. We can identify two contributions to the generalized force:
\begin{equation}
   \braket{\Psi_0|\hat{K}_\theta|\Psi_0}=\frac{1}{2\pi}\int_0^\infty\braket{\psi_E|\left[\hat{a}^\dagger\right]^\top\mathbf{Q}_\theta\left[\hat{a}\right]|\psi_E}\left|c\left(E\right)\right|^2\mathrm{d}E+\frac{1}{4\pi}\int_0^\infty\tr\left(\mathbf{Q}_\theta\right)\mathrm{d}E\,.\label{eq:K_manip_plus_vac}
\end{equation}
The second term, which is independent of the input state, is solely due to the vacuum fluctuations of the electromagnetic field, as is discussed in more detail in Section \ref{sec:VacForces}. The first term, on the other hand, can be engineered by proper choice of the input state. One is free to select a single operating frequency or choose a frequency window at which one desires to perform micromanipulation.

In the following, we derive the optimal input states at a single fixed frequency. As we will see below, a general feature of these optimal quantum input states is that their spatial profiles are always the classical ones, i.e., those obtained by an eigenvalue decomposition of the classical GWS matrix. Examples for such optimal spatial profiles are shown in Fig.\ \ref{fig:Optimal_micromanipulation}. Moreover, the choice of the quantum state that is injected into this classical channel does not change the resulting mean generalized force, as long as the quantum states carry the same mean total photon number $\nu$.

We write the QWS operator from Eq.\ (\ref{eq:QWS_GWS}) in the $\mathcal{Q}$ representation using Eqs.\ (\ref{eq:Q_eigdecomp}) and (\ref{eq:Annih_b_def}):
\begin{equation}
    \hat{Q}_\theta=\sum_{i=1}^N\hat{\nu}_i\lambda_i+\frac{1}{2}\sum_{i=1}^N\lambda_i\,,\label{eq:QWS_Qrepr}
\end{equation}
where $\hat{\nu}_i:=\hat{b}_i^\dagger\hat{b}_i$ is the photon number operator in the $i^\mathrm{th}$ eigenchannel of $\mathbf{Q}_\theta$. With the respective mean photon numbers $\nu_i:=\braket{\psi|\hat{\nu}_i|\psi}\geq 0$, we can write the expectation value of the generalized force as
\begin{equation}
    \braket{\psi|\hat{Q}_\theta|\psi}=\sum_{i=1}^N\nu_i\lambda_i+\frac{1}{2}\sum_{i=1}^N\lambda_i\,.
\end{equation}
Each photon in channel $i$ deposits a generalized force of value $\lambda_i$ onto the target. The mean force is additive in the mean photon number and correlations between photons are irrelevant in this case (similar to the radiation pressure force in cavity optomechanics \cite{Aspelmeyer2014}). The optimal way of using all $\nu=\sum_{i=1}^N\nu_i$ photons is to put them all into the channel corresponding to the highest eigenvalue $\lambda_{i_\mathrm{max}}$. This optimal state has a well-defined spatial shape which matches the classical optimum, see also Fig.\ \ref{fig:Optimal_micromanipulation}. The resulting optimal expectation value of the generalized force is
\begin{equation}
    \max_{\ket{\psi},\braket{\psi|\hat{\nu}|\psi}=\nu}\braket{\psi|\hat{Q}_\theta|\psi}=\nu\lambda_{i_\mathrm{max}}+\frac{1}{2}\sum_{i=1}^N\lambda_i\,.\label{eq:max_Q}
\end{equation}
This solution is specified just by the mean photon numbers $\nu_i$ in the GWS eigenchannels. This means that there is a degeneracy regarding the optimal input state: Whereas injecting all photonic resources into the channel $i_\mathrm{max}$ is sufficient for reaching optimality, the specific type of quantum state that is injected into this channel is irrelevant.

The considerations above focus on the mean force only. For precise nanoscale micromanipulation, however, also the fluctuations of the force must be minimized. These fluctuations are measured by the standard deviation of the generalized force operator $\hat{K}_\theta$. Here, we may exploit the degeneracy of the optimal input state mentioned above. For simplicity, we make the approximation that contributions from different parts of the frequency spectrum are independent of each other (although they are known to exist \cite{Mosk2012,Ambichl2017_Super}). This way, the variance of $\hat{K}_\theta$ is just the integrated variance of the QWS operator $\hat{Q}_\theta$. The previously mentioned degeneracy gives us room to choose the parameters of the input state in such a way as to minimize the standard deviation of $\hat{Q}_\theta$ while keeping its expectation value constant. Here, as an example, we focus on Gaussian states as they can readily be prepared in experiments and they allow for an analytical theoretical treatment \cite{Lassen2007,Kronwald2014,Andersen2016,Schnabel2017}. It turns out that for any fixed mean photon number $\nu$ there is a nontrivial mixture of mean coherent amplitude and squeezing that results in a minimal standard deviation of $\hat{Q}_\theta$. The details are provided in Appendix \ref{sec:App_Micromanip_MinVar}. Figure \ref{fig:OptMM_pbeta} shows the optimal mean coherent amplitude and the optimal squeezing strength, both in the relevant channel $i_\mathrm{max}$, as a function of $\nu$. The squeezing direction is always parallel to the coherent amplitude. The advantage one gains from this strategy is significant: When compared to the optimal classical (``unsqueezed'') state, one is able to reduce the standard deviation by more than half beyond a mean photon number of $\nu=49$ (see Fig.\ \ref{fig:OptMM_gain}), even though the necessary squeezing strength is not exceptionally high ($p_\mathrm{opt}\approx\SI{7.65}{\decibel}$ for $\nu=49$), increasing at most logarithmically with $\nu$,
\begin{equation}
    p_\mathrm{opt}\left(\nu\right)=\frac{1}{6}\ln\left(4\nu\right)+O\left(\frac{1}{\nu}\right)\,.
\end{equation}
These results hold for first maximizing the expectation value of $\hat{Q}_\theta$ and then minimizing its variance. The order of maximization and minimization is crucial here. Likewise, one can also consider a joint optimization of both quantities with appropriately weighted penalties.

\begin{figure*}[t!]
    \centering
    \subcaptionbox{Optimal Gaussian input state\label{fig:OptMM_pbeta}}{
        \includegraphics[scale=0.86]{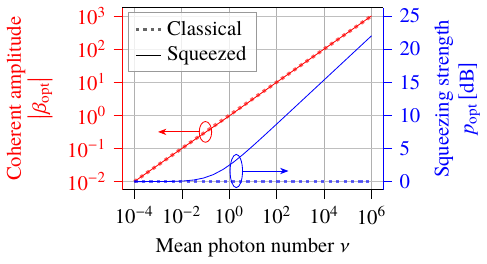}
    }
    \hfill
    \subcaptionbox{Reduction factor of standard deviation (std)\label{fig:OptMM_gain}}{
        \includegraphics[scale=0.86]{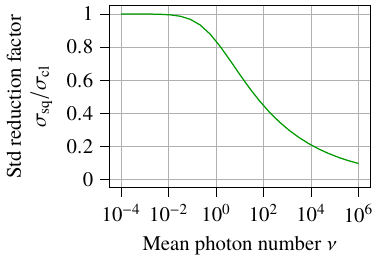}
    }
    \caption{Quantum-enhanced optimal micromanipulation. The strength of fluctuations in the opto-mechanical force is given by the standard deviation $\sigma$ of the corresponding QWS operator. The aim is to minimize those fluctuations while keeping the mean force (i.e., the expectation value of the QWS operator) constant. We consider Gaussian input states and compare the optimal squeezed state to the optimal classical (i.e., coherent) state. \textbf{(a)} Solid lines show the parameters for the optimal squeezed state as a function of the mean photon number $\nu$. The optimal absolute value of the mean coherent amplitude and the optimal squeezing strength are denoted by $\left|\beta_\mathrm{opt}\right|$ and $p_\mathrm{opt}$, respectively. Mind the logarithmic scale for $\nu$ and $\left|\beta_\mathrm{opt}\right|$, whereas $p_\mathrm{opt}$ is shown on a linear scale. For comparison, the best classical state, characterized by $p_\mathrm{opt}=0$ and $\left|\beta_\mathrm{opt}\right|=\sqrt{\nu}$, is indicated by dotted lines---the difference between the solid red line and the dotted red line is minute. \textbf{(b)} This plot shows the factor by which the force fluctuations are reduced when using the optimal squeezed state ($\sigma_\mathrm{sq}$) instead of the optimal classical state ($\sigma_\mathrm{cl}$).}\label{fig:OptMM}
\end{figure*}

\section{Vacuum forces\label{sec:VacForces}}

The quantum Wigner-Smith operator $\hat{Q}_\theta$ describes the forces of the radiation field upon the parameter $\theta$---forces that can be harnessed for optical micromanipulation. In this section, we aim to discuss the physical meaning of the trace term appearing in Eq.\ (\ref{eq:QWS_GWS}). Note that this term is the vacuum expectation value of $\hat{Q}_\theta$:
\begin{equation}
    \braket{\mathbf{0}|\hat{Q}_\theta|\mathbf{0}}=\frac{1}{2}\tr\left(\mathbf{Q}_\theta\right)\,.
\end{equation}
This trace term has a physical meaning with a distinguished history: it describes the forces of the quantum vacuum \cite{Scheel2008,Rodriguez2011,Buhmann2012,SimpsonLeonhardt2015}. These are the generalized van der Waals forces \cite{Dzyaloshinskii1961,Buhmann2012}---the Casimir-Polder forces between two molecules \cite{CasimirPolder1948} or the Casimir forces \cite{Casimir1948} between two or more dielectric bodies. Or, as $\theta$ can be rather general, these are the torques between birefringent plates \cite{Somers2018}, or the capillary forces \cite{Buhmann2012} that lift up water to the leaves of trees (limiting the maximal height to which trees can grow \cite{Koch2004}).

First, let us give an intuitive heuristic explanation of the nature of vacuum forces. Zero-point fluctuations of the electromagnetic field manifest themselves as omnipresent virtual photons that equally populate all modes at all frequencies. For a specific frequency, this leads to a (nonvirtual, i.e.\ real) force which is an equally weighted sum over all contributions from all the modes. Mathematically, this is expressed by the trace
\begin{equation}
    \tr\left(\mathbf{Q}_\theta\right)=\sum_{m=1}^N\mathbf{e}_m^\dagger\mathbf{Q}_\theta\mathbf{e}_m\,.
\end{equation}
Here, $\mathbf{e}_m$ are the unit basis vectors describing the modes of the electromagnetic field. The consequence of this trace term is that a finite force is transmitted onto a target even when no light is injected into the system at all.

To see on a more technical level that the trace of the GWS matrix $\mathbf{Q}_\theta$ describes the vacuum forces, we derive the latter from first principles along the same lines as the established literature on the relationship between scattering theory and vacuum forces \cite{Lambrecht2006,Wirzba2008}. Consider the scattering phase $\eta$, defined as the sum of all eigenphases $\eta_k$ of the unitary scattering matrix $\mathbf{S}$ with eigenvalues $\mathrm{e}^{\mathrm{i}\eta_k}$:
\begin{equation}
    \eta=\sum_{k=1}^N\eta_k = -\mathrm{i}\ln\left(\det\left(\mathbf{S}\right)\right)\,.\label{eq:scatteringphase}
\end{equation}
The scattering phase, in turn, provides direct access to the density of states $\rho\left(E\right)$ according to Krein's trace formula \cite{Birman1962,Faulkner1977},
\begin{equation}
    \rho\left(E\right)=\rho_0\left(E\right)+\frac{1}{2\pi}\frac{\partial\eta}{\partial E}\,,\label{eq:dos}
\end{equation}
where $\rho_0\left(E\right)$ is the density of states for free space which is independent of $\theta$. Since 
\begin{equation}
    \frac{\partial\eta}{\partial\theta}=-\mathrm{i}\tr\left(\mathbf{S}^\dagger\frac{\partial\mathbf{S}}{\partial\theta}\right) = \tr\left(\mathbf{Q}_\theta\right)\,,\label{eq:dphi_dtheta}
\end{equation}
for arbitrary $\theta$ (including $E$), we have:
\begin{equation}
    \rho\left(E\right)=\rho_0\left(E\right)+\frac{1}{2\pi}\tr\left(\mathbf{Q}_E\right)\,.
\end{equation}
Now, the vacuum force $K_\theta^\mathrm{vac}$ upon $\theta$ is the negative derivative of the vacuum energy with respect to $\theta$:
\begin{equation}
    K_\theta^\mathrm{vac}=-\frac{\partial}{\partial\theta}\int_0^\infty\frac{E}{2}\rho({E})\mathrm{d}E=-\frac{1}{4\pi}\int_0^\infty E\frac{\partial^2\eta}{\partial E\,\partial\theta}\,\mathrm{d}E\,.
\end{equation}
Integrating by parts and using Eq.\ (\ref{eq:dphi_dtheta}) gives 
\begin{equation}
    K_\theta^\mathrm{vac}=\frac{1}{4\pi}\int_0^\infty\tr\left(\mathbf{Q}_\theta\right)\mathrm{d}E=\braket{0|\hat{K}_\theta|0}\,.\label{eq:K_theta_vac}
\end{equation}
This formula relates the classical GWS matrix $\mathbf{Q}_\theta$ to the vacuum force and agrees with our finding in Eq.\ (\ref{eq:K_manip_plus_vac}). In deriving it by partial integration, we assumed that $\tr\left(\mathbf{Q}_\theta\right)$ vanishes for $E\rightarrow\infty$ (or is infinitely oscillatory such that it vanishes effectively). This assumption is based on the physical fact that the vacuum forces \cite{SimpsonLeonhardt2015} originate from reflections between scatterers and that those reflections vanish for $E\rightarrow\infty$ due to dispersion \cite{LandauLifshitz5_1980}.

The bare vacuum energy and its density is infinite, but the part of the energy that can do physical work is finite. Renormalization---the subtraction of the infinite, unphysical contribution from the vacuum energy---is required. Physically motivated renormalization methods have been suggested right from the beginning of Casimir physics research in the late 1940's. Casimir himself \cite{Casimir1948} extracted the part of the vacuum energy that can do physical work by taking the difference between a finite and an infinite cavity. Taking the difference between vacuum energies for finite and infinite distances is also the basis for renormalization in modern numerical methods for calculating the Casimir force between arbitrary dielectric bodies \cite{Reid2009}. But this renormalization method cannot determine the Casimir force of the dielectric upon itself, in particular in inhomogeneous media \cite{Simpson2013,Griniasty2017,Efrat2021}, because one cannot take such media apart to infinity for determining their intrinsic vacuum stresses. Ref.\ \cite{Simpson2013} shows that for inhomogeneous media, the simple ansatz of discretizing such media into small homogeneous sections does not converge in the continuum limit. The QWS operator may serve as a starting point to overcome these problems and to provide an understanding of the physical phenomena that underlie mathematical renormalization procedures \cite{Hoerner2023}.

\section{Quantum metrology\label{sec:QuMetrology}}

\subsection{Quantum Fisher information}

In this section, we interpret $\theta$ as a parameter of the scattering system, the value of which we want to estimate. In order to arrive at an estimate of $\theta$ with as little uncertainty as possible, quantum metrology addresses not only the question of how to prepare a corresponding probe state, but also how to make optimal measurements on the transformed probe state and how to process the data collected in the measurements \cite{Helstrom1969,Braunstein1994,Holevo2011,Giovannetti2011,Polino2020}. A well-known result in this context is that the measurement uncertainty is always larger than the inverse of the quantum Fisher information (QFI) $F_\theta$, as is expressed by the quantum Cram\'{e}r-Rao bound
\begin{equation}
    \mathbb{V}\left[\tilde{\theta}\left(X\right)\right]\geq\frac{1}{MF_\theta}\,,\label{eq:Qu_CramerRao}
\end{equation}
where $\tilde{\theta}\left(X\right)$ is an unbiased estimator for $\theta$ based on the measurement outcome $X$ ($\mathbb{V}$ denotes the statistical variance and $M$ is the number of repeated independent measurements). The QFI is only determined by the $\theta$-dependence of the quantum state that interacted with the system, but it is independent of the measurement scheme and independent of the estimator.

Minimizing the uncertainty therefore requires maximizing the QFI with respect to the probe state, resulting in a minimal right hand side for the quantum Cram\'{e}r-Rao bound in Eq.\ (\ref{eq:Qu_CramerRao}). We neither treat the measurement (i.e.\ the choice of $X$) nor the estimation procedure (i.e.\ the choice of the estimator $\tilde{\theta}\left(X\right)$). Given a specific probe state, it is always possible to saturate the quantum Cram\'{e}r-Rao bound by choosing an appropriate measurement \cite{Braunstein1994,Polino2020} and (in the asymptotic limit of many measurements) a maximum likelihood estimator \cite{VanTrees2013}. If experimental or other restrictions inhibit the implementation of the theoretically optimal measurement, then one has to resort to maximizing the classical Fisher information, which depends on the probe state (like the QFI does), but additionally also on the measurement scheme. In this optimization problem one can then impose appropriate restrictions regarding the probe state and the measurement scheme or even fix the measurement altogether, e.g., to a homodyne detection.

Because the QFI is convex with respect to the probe state, the optimal state is pure, $\hat{\rho}_0=\ket{\psi_0}\bra{\psi_0}$ \cite{Fujiwara2001,Fiderer2019}. The scattering-induced transformation of the probe state is governed by the unitary operator $\hat{U}\left(\theta\right)$ defined in Eq.\ (\ref{eq:Bogoliubov_trafo}). In this case of a unitarily transformed pure state, the QFI can be easily expressed as the variance of the QWS operator with respect to the probe state \cite{Liu2020}:
\begin{equation}
    F_\theta=4\mathbb{V}_{\ket{\psi_0}}\left[\hat{Q}_\theta\right]=4\left(\braket{\psi_0|\hat{Q}_\theta^2|\psi_0}-\braket{\psi_0|\hat{Q}_\theta|\psi_0}^2\right)\,.\label{eq:QFI_VarQ}
\end{equation}

Moreover, combining Eqs.\ (\ref{eq:Qu_CramerRao}) and (\ref{eq:QFI_VarQ}) for a single measurement $M=1$, one obtains a fundamental uncertainty principle \cite{Braunstein1994}
\begin{equation}
    \mathbb{V}\left[\tilde{\theta}\left(X\right)\right]\mathbb{V}_{\ket{\psi_0}}\left[\hat{Q}_\theta\right]\geq\frac{1}{4}\,,\label{eq:uncertainty_principle}
\end{equation}
which is more general than the standard uncertainty principle because $\theta$ is not restricted to being a quantum operator---indeed it can be any parameter of the system. The inequality (\ref{eq:uncertainty_principle}) tells us, in the language of scattering matrices, that on a fundamental level, gaining more information about a physical parameter $\theta$ comes at the cost of causing greater perturbations in $\theta$.

\subsection{Gaussian probe states}

Using these well-known results from quantum metrology \cite{Braunstein1994,Liu2020}, we can now connect the QFI to the classical scattering matrix. For monochromatic classical light which is described by coherent quantum states $\ket{\bm{\alpha}}^{\mathcal{M}}$, $\bm{\alpha}\in\mathbb{C}^N$ being the amplitudes of the $N$ input modes (see Subsection \ref{subsec:Fund_quant_optics}), we obtain (see \cite{Bouchet2021} and Appendix \ref{sec:App_QFI_Gaussian})
\begin{equation}
    F_\theta=4\bm{\alpha}^\dagger\mathbf{Q}_\theta^2\bm{\alpha}\,.\label{eq:VarQ_Coherent}
\end{equation}
Given a specific mean total photon number $\nu$, which is equal to $\left\Vert\bm{\alpha}\right\Vert^2$ and proportional to the energy of the light (apart from the zero point energy), the optimal metrological input state that maximizes the QFI in Eq.\ (\ref{eq:VarQ_Coherent}) is given by the eigenvector of the GWS matrix $\mathbf{Q}_\theta$ that corresponds to the eigenvalue $\lambda_{i_{\mathrm{hav}}}$ with the \textbf{h}ighest \textbf{a}bsolute \textbf{v}alue. The maximized QFI itself is given by
\begin{equation}
    F_\theta=4\lambda_{i_{\mathrm{hav}}}^2\nu\,.\label{eq:QFI_SQL}
\end{equation}

\begin{figure}[t!]
    \centering
    \includegraphics[scale=0.9]{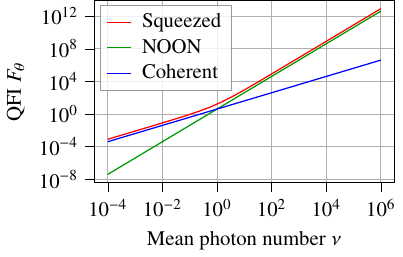}
    \caption{QFI $F_\theta$ as a function of the mean total photon number $\nu$ for different choices for the probe state in unitary quantum metrology. Here, we assume that the GWS matrix has a generic eigenvalue spectrum corresponding to a uniform distribution $\mathcal{U}\left(-1,1\right)$. The optimal coherent probe state (blue line) exhibits the scaling $F_\theta\propto\nu$, indicating the SQL. Both the optimal photon number state, which is a NOON state (green line), and the optimal squeezed state (red line) are able to reach the HL, as indicated by the scaling $F_\theta\propto\nu^2$, for large values of $\nu$.}
    \label{fig:QFI_scalings}
\end{figure}

Let us now consider another case of experimental relevance \cite{Lassen2007,Kronwald2014,Andersen2016,Schnabel2017}: monochromatic squeezed states. It is most convenient to express the corresponding QFI in the $\mathcal{Q}$ representation, see Eq.\ (\ref{eq:QFI_Gaussian}) in Appendix \ref{sec:App_QFI_Gaussian}. The maximization of the QFI under the constraint of a given mean total photon number $\nu$ is carried out in Appendix \ref{sec:App_QFI_OptGaussian}. The resulting optimal Gaussian probe state is characterized as follows: All channels are populated by the vacuum state and all resources (in terms of the mean photon number or, equivalently, the energy of the light) are used to squeeze the vacuum in the channel corresponding to the eigenvalue of the GWS matrix $\mathbf{Q}_\theta$ with the highest absolute value. Since the coherent amplitude $\bm{\beta}$ vanishes in this case, there is no preferred direction in the photonic quantum phase space, which is why the squeezing angle is irrelevant here. The corresponding QFI is given by
\begin{equation}
    F_\theta=8\lambda_{i_{\mathrm{hav}}}^2\nu\left(\nu+1\right)\,,\label{eq:QFI_HL}
\end{equation}
which is (for $\nu>0$) always strictly greater than the QFI (\ref{eq:QFI_SQL}) of the optimal coherent probe state with the same number of photons, see also Fig.\ \ref{fig:QFI_scalings}.

Another important difference between the last two equations is the dissimilar scaling of the QFI $F_\theta$ with respect to the mean total photon number $\nu$. The physical reason behind this observation is that photons in a coherent state are uncorrelated. This necessarily limits the estimation precision to the so-called standard quantum limit, indicated by the linear scaling $F_\theta\propto\nu$ in Eq.\ (\ref{eq:QFI_SQL}), see also Fig.\ \ref{fig:QFI_scalings}. In order to surpass this limit and attain what is known as the Heisenberg limit, characterized by the quadratic scaling $F_\theta\propto\nu^2$ for $\nu\gg 1$ in Eq.\ (\ref{eq:QFI_HL}), one must resort to quantum correlated, i.e., entangled photons \cite{Holland1993,Giovannetti2004,Giovannetti2011,Polino2020,Treps2002}. Such quantum correlations are provided by squeezing \cite{Paris1999,VanLoock2000,Wolf2003,Silberhorn2001,Korolkova2002}, see also Fig.\ \ref{fig:QFI_scalings}.

What the optimal coherent probe state and the optimal Gaussian probe state have in common is that only the channel $i_\mathrm{hav}$ is populated by photons. For the waveguide setup introduced in Subsection \ref{subsec:Scat_cl_optics}, the spatial structure of such a channel is shown in Fig.\ \ref{fig:mm_h_max} for $\theta=x$ and Fig.\ \ref{fig:mm_v_min} for $\theta=y$.

\subsection{Photon number probe states\label{subsec:Ph_num_probe}}

As a last example, let us now search for optimized probe states, which are characterized by a well-defined number of photons, $\nu\in\mathbb{N}_0$. The space of such ``photon number states'' is spanned by Fock states, which form an orthonormal basis. Each Fock state is characterized by the number of photons $\nu_i\in\mathbb{N}_0$ in each channel $i$, such that $\sum_{i=1}^N\nu_i=\nu$. Here, we work with the Fock states associated with the $\mathcal{Q}$ representation. Throughout the rest of this subsection, we stay in the $\mathcal{Q}$ representation and hence suppress the corresponding label. We denote the Fock states with the symbol $\ket{\bm{\nu}}$, where the $\nu_i$ are the components of the vector $\bm{\nu}$. The Fock states are created from the vacuum state $\ket{\mathbf{0}}$ by applying the appropriate combination of creation operators $\hat{b}_i^\dagger$,
\begin{equation}
    \ket{\bm{\nu}}:=\prod_{i=1}^N\frac{1}{\sqrt{\nu_i!}}\left(\hat{b}_i^\dagger\right)^{\nu_i}\ket{\mathbf{0}}\,.
\end{equation}
We denote the set of all $N$-mode Fock states with a total number of $\nu$ photons with
\begin{equation}
    \mathcal{F}_\nu^N:=\left\{\bm{\nu}\in\mathbb{N}_0^N:\sum_{i=1}^N\nu_i=\nu\right\}\,.
\end{equation}
A photon number state is a superposition of those states,
\begin{equation}
    \ket{\psi}=\sum_{\bm{\nu}\in\mathcal{F}_\nu^N}\braket{\bm{\nu}|\psi}\ket{\bm{\nu}}=:\sum_{\bm{\nu}\in\mathcal{F}_\nu^N}\psi_{\bm{\nu}}\ket{\bm{\nu}}\,.\label{eq:Fock_superposition}
\end{equation}

Recalling the photon number operators $\hat{\nu}_i=\hat{b}_i^\dagger\hat{b}_i$ and $\hat{\nu}=\sum_{i=1}^N\hat{\nu}_i$, the most important mathematical properties of the Fock states are ($\mathbf{e}_i$ denotes the $i^\text{th}$ unit vector)
\begin{align}
    \hat{b}_i\ket{\bm{\nu}}&=\sqrt{\nu_i}\ket{\bm{\nu}-\mathbf{e}_i}\,,\\
    \hat{b}_i^\dagger\ket{\bm{\nu}}&=\sqrt{\nu_i+1}\ket{\bm{\nu}+\mathbf{e}_i}\,,\\
    \hat{\nu}_i\ket{\bm{\nu}}&=\nu_i\ket{\bm{\nu}}\,,\\
    \hat{\nu}\ket{\bm{\nu}}&=\sum_{i=1}^N\nu_i\ket{\bm{\nu}}\,.
\end{align}
Fock states are eigenstates of the photon number operators. Since the QWS operator (\ref{eq:QWS_Qrepr}) commutes with all photon number operators $\hat{\nu}_i$, its eigenstates are also Fock states (in the $\mathcal{Q}$ representation). Focusing on the ``operator-valued'' part of the QWS operator $\hat{Q}_\theta^\mathrm{I}$ in Eq.\ (\ref{eq:Q_theta_I}), we obtain
\begin{equation}
    \hat{Q}_\theta^\mathrm{I}\ket{\bm{\nu}}=\sum_{i=1}^N\lambda_i\hat{\nu}_i\ket{\bm{\nu}}=\sum_{i=1}^N\lambda_i\nu_i\ket{\bm{\nu}}=:\lambda_{\bm{\nu}}\ket{\bm{\nu}}\,.\label{eq:QWS_eigen}
\end{equation}

With this insight, we can immediately write (see Eqs.\ (\ref{eq:Fock_superposition}) and (\ref{eq:QWS_eigen}))
\begin{equation}
    \mathbb{V}_{\ket{\psi}}\left[\hat{Q}_\theta\right]=\sum_{\bm{\nu}\in\mathcal{F}_\nu^N}\left|\psi_{\bm{\nu}}\right|^2\lambda_{\bm{\nu}}^2-\left(\sum_{\bm{\nu}\in\mathcal{F}_\nu^N}\left|\psi_{\bm{\nu}}\right|^2\lambda_{\bm{\nu}}\right)^2\,.\label{eq:Var_Q_Fock}
\end{equation}
This corresponds to the variance of the discrete distribution of values $\lambda_{\bm{\nu}}$ with respective probabilities $\left|\psi_{\bm{\nu}}\right|^2$. Before we proceed, we sort the eigenvalues $\lambda_i$ in descending order such that $\lambda_{i_\mathrm{max}}=\lambda_1\geq\lambda_2\geq\ldots\geq\lambda_N=\lambda_{i_\mathrm{min}}$. To find the optimal probe state which maximizes Eq.\ (\ref{eq:Var_Q_Fock}), we invoke Popoviciu's inequality on variances \cite{Popoviciu1935}, which states that the variance of such a probability distribution $p\left(\lambda_{\bm{\nu}}\right)$ is bound from above according to
\begin{equation}
    \mathbb{V}_{\ket{\psi}}\left[\hat{Q}_\theta\right]\leq\frac{\left((\lambda_{\bm{\nu}})_\mathrm{max}-(\lambda_{\bm{\nu}})_\mathrm{min}\right)^2}{4}=\nu^2\frac{\left(\lambda_1-\lambda_N\right)^2}{4}\,.\label{eq:Popoviciu}
\end{equation}
In the last step we used the fact that $(\lambda_{\bm{\nu}})_\mathrm{max}=\nu\lambda_1$ and likewise for the minimum, see also Eq.\ (\ref{eq:QWS_eigen}).

The inequality (\ref{eq:Popoviciu}) is saturated, i.e.\ the variance (\ref{eq:Var_Q_Fock}) and thus the QFI (see Eq.\ (\ref{eq:QFI_VarQ})) is maximized, for
\begin{equation}
    \left|\psi_{\bm{\nu}}\right|^2_\mathrm{opt}=\begin{cases}
        \frac{1}{2} & \bm{\nu}\in\left\{\nu\mathbf{e}_1,\nu\mathbf{e}_N\right\}\\
        0 & \text{else},
    \end{cases}
\end{equation}
yielding the optimal probe state
\begin{equation}
    \ket{\psi_\mathrm{opt}}=\frac{1}{\sqrt{2}}\left(\mathrm{e}^{\mathrm{i}\varphi_1}\ket{\nu\mathbf{e}_1}+\mathrm{e}^{\mathrm{i}\varphi_N}\ket{\nu\mathbf{e}_N}\right)\,.\label{eq:NOON}
\end{equation}
The phase factors $\mathrm{e}^{\mathrm{i}\varphi_1}$ and $\mathrm{e}^{\mathrm{i}\varphi_N}$ are arbitrary. This is a so-called NOON state \cite{Boto2000,Lee2002}, establishing a maximal degree of entanglement between the two channels corresponding to the largest and smallest eigenvalue of the GWS matrix. Such superpositions of spatial modes have been created experimentally to increase measurement sensitivities \cite{Hiekkamaeki2021_spatial,Hiekkamaeki2021_angular}. The QFI with respect to the NOON state (\ref{eq:NOON}) is
\begin{equation}
    F_\theta=\left(\lambda_1-\lambda_N\right)^2\nu^2\,.
\end{equation}
The quadratic scaling $F_\theta\propto\nu^2$ indicates that the Heisenberg limit is reached using this optimal probe state, see also Fig.\ \ref{fig:QFI_scalings}. However, we find that, regarding the QFI, the optimal Gaussian probe state with the same mean total photon number $\nu$, see Eq.\ (\ref{eq:QFI_HL}), surpasses the NOON state by a factor of at least 2:
\begin{equation}
    F_\theta^\text{optimal Gauss}\geq 2F_\theta^\text{optimal NOON}\,.
\end{equation}

\begin{figure}[t!]
    \centering
    \subcaptionbox{Horizontal displacement\label{fig:NOON_hori}}{
        \includegraphics[scale=0.88]{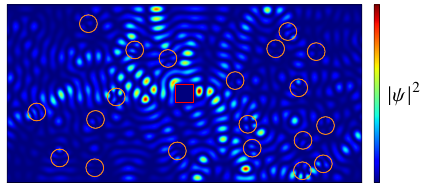}
    }
    \hfill
    \subcaptionbox{Vertical displacement\label{fig:NOON_vert}}{
        \includegraphics[scale=0.88]{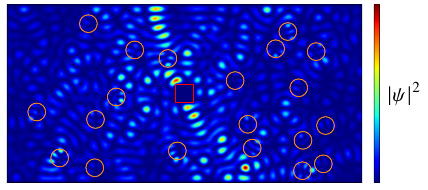}
    }
    \caption{Spatial probability densities of single photon NOON states in the system described in Subsection \ref{subsec:Scat_cl_optics} for different choices of the parameter $\theta$. \textbf{(a)} The parameter $\theta$ is taken as the horizontal position $x$ of the target (red square). \textbf{(b)} Here, $\theta$ is taken as the vertical position $y$ of the target (red square).}
    \label{fig:NOON}
\end{figure}

For a concrete demonstration of NOON states, we consider the waveguide setup introduced in Subsection \ref{subsec:Scat_cl_optics}. For a single photon ($\nu=1$), the spatial structures of the probability densities of two selected NOON states are illustrated in Fig.\ \ref{fig:NOON}. The plots make it clear that these NOON states build up intensity right where the target changes when $\theta$ is varied (in positive or negative direction). To make this more specific, we see, e.g., that for $\theta=x$ the NOON state builds up intensity to the left and the right of the target (see Fig.\ \ref{fig:NOON_hori}). In contrast, eigenstates of the corresponding GWS matrix $\mathbf{Q}_x$ lead to a high intensity only on one side of the target (see Figs.\ \ref{fig:mm_h_max} and \ref{fig:mm_h_min}).

\section{Summary\label{sec:Summary}}

We discuss here a simple formalism to describe the interaction between the spatial as well as the quantum degrees of freedom of light and a local parameter of a linear, but otherwise arbitrarily complex scattering medium. This formalism explicitly connects quantum micromanipulation, vacuum forces and quantum metrology with classical scattering matrices, which are experimentally measurable in a noninvasive manner. We show how to design protocols for optimal micromanipulation as well as for optimal parameter estimation by shaping both the spatial and the quantum degrees of freedom of light. An important result of this analysis is that for micromanipulation, the spatial shapes of the optimal classical fields are also the optimal quantum ones. Engineering their quantum properties allows one, however, to reduce the noise in these fields. The discussed framework treats radiation pressure and the Casimir effect on the same footing, which enables the use of light fields to compensate for vacuum forces. In the context of quantum metrology, we show that both NOON states and squeezed Gaussian states can reach the Heisenberg limit in complex scattering media. The relevant channels are given by eigenvectors of the classical GWS matrix. This matrix is derived from the system's classical scattering matrix, which is especially convenient to describe systems with many spatial modes such as complex scattering media. The presented framework should also be extendable to systems with loss or incomplete channel control. We hope that our work will stimulate exchange between the different communities addressed in this tutorial.

\begin{backmatter}

\bmsection{Funding}
L.M.R.\ and S.R.\ were supported by the Austrian Science Fund (FWF) through Project No.\ P32300 (WAVELAND). U.L.\ was supported by the Israel Science Foundation and the Murray B.\ Koffler Professorial Chair.

\bmsection{Acknowledgment}
The authors thank Sylvain Gigan, Michael Horodynski, Ivor Kre\v{s}i\'{c}, Matthias K{\"u}hmayer, Allard P.\ Mosk, Nicolas Treps and Matthias Zens for insightful and stimulating discussions.

\bmsection{Disclosures.}
The authors declare no conflicts of interest.

\bmsection{Data availability}
The data that support the plots within this paper are available from the corresponding author upon reasonable request.


\end{backmatter}

\appendix

\section{Proof of Eq.\ (\ref{eq:QWS_GWS})\label{sec:App_proof_QWS_GWS}}

\subsection{First lemma}

Let $Y\left(\theta\right)$ be a $\theta$-dependent matrix or operator, then
\begin{align}
    \mathrm{e}^{-Y\left(\theta\right)}\frac{\partial\mathrm{e}^{Y\left(\theta\right)}}{\partial\theta}&=\int_0^1\mathrm{e}^{-tY\left(\theta\right)}Y'\left(\theta\right)\mathrm{e}^{tY\left(\theta\right)}\mathrm{d}t\label{eq:Deriv_exp_int}\\
    &=\sum_{r=0}^\infty\frac{\left(-1\right)^r}{\left(r+1\right)!}\left[Y\left(\theta\right),Y'\left(\theta\right)\right]_r,\label{eq:Deriv_exp_sum}
\end{align}
where $\left[\cdot,\cdot\right]_r$ is the $r$-fold nested commutator. Eq.\ (\ref{eq:Deriv_exp_int}) is derived in Ref.\ \cite{Snider1964} (Appendix B) and Eq.\ (\ref{eq:Deriv_exp_sum}) is obtained using the Hadamard lemma
\begin{equation}
    \mathrm{e}^A B\mathrm{e}^{-A}=\sum_{r=0}^\infty\frac{1}{r!}\left[A,B\right]_r.
\end{equation}

\subsection{Second lemma}

The second lemma we need for our proof is that for all $\mathbf{J},\mathbf{K}\in\mathbb{C}^{N\times N}$:
\begin{equation}
    \left[\hat{J},\hat{K}\right]=\hat{C},\label{eq:Lemma_2_Commutators}
\end{equation}
where $\mathbf{C}:=\left[\mathbf{J},\mathbf{K}\right]$ is the matrix commutator and each matrix $\mathbf{M}\in\{\mathbf{J},\mathbf{K},\mathbf{C}\}$ gets mapped to the operator $\hat{M}:=\left[\hat{a}^\dagger\right]^\top\mathbf{M}\left[\hat{a}\right]\in\{\hat{J},\hat{K},\hat{C}\}$. In other words, the commutation relations of the bosonic creation and annihilation operators correctly encode the ordinary matrix commutation rules.

This lemma is proven using the fundamental bosonic commutation relations. A corollary of this lemma is that the same relation also holds for the nested commutators:
\begin{equation}
    \left[\hat{J},\hat{K}\right]_{r}=\widehat{C_r}.\label{eq:Second_lemma}
\end{equation}

\subsection{The proof}

If $\mathbf{S}\left(\theta\right)$ is a unitary scattering matrix, then
\begin{equation}
    \mathbf{L}\left(\theta\right):=-\mathrm{i}\ln\left(\mathbf{S}\left(\theta\right)\right)
\end{equation}
is a Hermitian matrix. The generalized Wigner-Smith (GWS) matrix is obtained as (see Eqs.\ (\ref{eq:GWS_def}), (\ref{eq:Deriv_exp_int}) and (\ref{eq:Deriv_exp_sum}))
\begin{align}
    \mathbf{Q}_\theta&=\int_0^1\mathrm{e}^{-\mathrm{i}t\mathbf{L}\left(\theta\right)}\mathbf{L}'\left(\theta\right)\mathrm{e}^{\mathrm{i}t\mathbf{L}\left(\theta\right)}\mathrm{d}t\label{eq:GWS_proof_int}\\
    &=\sum_{r=0}^\infty\frac{\left(-\mathrm{i}\right)^r}{\left(r+1\right)!}\left[\mathbf{L}\left(\theta\right),\mathbf{L}'\left(\theta\right)\right]_r.\label{eq:GWS_proof_sum}
\end{align}
Note that the GWS matrix is \emph{not} just the derivative of the logarithm of the scattering matrix, i.e.\ in general $\mathbf{Q}_\theta\neq\mathbf{L}'\left(\theta\right)$.)

In order to calculate the quantum Wigner-Smith (QWS) operator, we first rewrite the corresponding transformation operator $\hat{U}\left(\theta\right)$ into normal order:
\begin{equation}
    \hat{U}\left(\theta\right)=\mathrm{e}^{\mathrm{i}\left[\hat{a}^\dagger\right]^\top\mathbf{L}\left(\theta\right)\left[\hat{a}\right]}\mathrm{e}^{\frac{\mathrm{i}}{2}\tr\left(\mathbf{L}\left(\theta\right)\right)}.\label{eq:U_normal_order}
\end{equation}

Using the product rule of differentiation, we can split the QWS operator $\hat{Q}_\theta=-\mathrm{i}\hat{U}^\dagger\left(\theta\right)\partial_\theta\hat{U}\left(\theta\right)$ into the sum of two expressions, namely
\begin{align}
    \hat{Q}_\theta^\mathrm{I}&:=-\mathrm{i}\mathrm{e}^{-\mathrm{i}\left[\hat{a}^\dagger\right]^\top\mathbf{L}\left(\theta\right)\left[\hat{a}\right]}\frac{\partial}{\partial\theta}\mathrm{e}^{\mathrm{i}\left[\hat{a}^\dagger\right]^\top\mathbf{L}\left(\theta\right)\left[\hat{a}\right]},\label{eq:QWS_1st_term}\\
    \hat{Q}_\theta^\mathrm{II}&:=-\mathrm{i}\mathrm{e}^{-\frac{\mathrm{i}}{2}\tr\left(\mathbf{L}\left(\theta\right)\right)}\frac{\partial}{\partial\theta}\mathrm{e}^{\frac{\mathrm{i}}{2}\tr\left(\mathbf{L}\left(\theta\right)\right)}.\label{eq:QWS_2nd_term}
\end{align}
The first term $\hat{Q}_\theta^\mathrm{I}$ has the form $\mathrm{e}^{-Y\left(\theta\right)}\partial_\theta\mathrm{e}^{Y\left(\theta\right)}$ so we can use the first lemma from above (Eq.\ (\ref{eq:Deriv_exp_sum})). Further employing the second lemma (Eq.\ (\ref{eq:Second_lemma})) and Eq.\ (\ref{eq:GWS_proof_sum}), it is straightforward to show that
\begin{equation}
    \hat{Q}_\theta^\mathrm{I}=\left[\hat{a}^\dagger\right]^\top\mathbf{Q}_\theta\left[\hat{a}\right].\label{eq:Q_theta_I}
\end{equation}
The second term $\hat{Q}_\theta^\mathrm{I}$ can be calculated using Eq.\ (\ref{eq:GWS_proof_int}):
\begin{equation}
    \hat{Q}_\theta^\mathrm{II}=\frac{1}{2}\tr\left(\mathbf{Q}_\theta\right).
\end{equation}

\section{Micromanipulation\label{sec:App_Micromanip}}

\subsection{Physical interpretation of the QWS operator\label{sec:App_Micromanip_PhysInterpret}}

Here, we want to establish a physical interpretation of the QWS operator $\hat{Q}_\theta$ as the quantum operator describing the generalized force conjugate to $\theta$. In general, a force is defined as the negative gradient of the Hamiltonian with respect to the parameter $\theta$. So in order to arrive at the desired correspondence, we have to express the ``scattering operator'' $\hat{U}$ from Eq.\ (\ref{eq:Bogoliubov_trafo}) in terms of the Hamiltonian $\hat{H}$. This is a well-known result from formal scattering theory, typically formulated in terms of a scattering matrix \cite{Verbaarschot1985}, but it holds equally for the quantum operator $\hat{U}$ following Refs.\ \cite{Fyodorov1997,Dittes2000},
\begin{align}
    \hat{U}&=\hat{1}-2\pi\mathrm{i}\hat{W}^\dagger\hat{G}\hat{W},\\
    \hat{G}&=\left(E-\hat{H}_\mathrm{eff}\right)^{-1},\\
    \hat{H}_\mathrm{eff}&=\hat{H}-\pi\mathrm{i}\hat{W}\hat{W}^\dagger,
\end{align}
where $\hat{G}$ is the Green's operator in the interior of the scattering region, $\hat{W}$ describes the coupling between the channel basis in the asymptotic region and the local basis at the boundary of the scattering region, $E$ is the energy and $\hat{H}_\mathrm{eff}$ is the effective Hamiltonian. In order to comprehend the connection to Ref.\ \cite{Dittes2000}, we rewrite $\hat{U}=(\hat{1}-\mathrm{i}\hat{K})(\hat{1}+\mathrm{i}\hat{K})^{-1}$ with $\hat{K}=\pi\hat{W}^\dagger (E-\hat{H})^{-1}\hat{W}$. This representation is identical to Eqs.\ (2.50) and (2.51) from \cite{Dittes2000} by virtue of identifying $\hat{W}=Q\mathscr{H}P$ and $\hat{H}=Q\mathscr{H}Q$, where $\mathscr{H}$ is the ``full'' Hamiltonian and $Q$ and $P$ are the projection operators onto the subspace of ``bound'' and ``scattering'' states, respectively.

Proceeding, we observe the following:
\begin{align}
    \hat{1}&=\left(E-\hat{H}_\mathrm{eff}\right)\hat{G}\\
    \implies \hat{G}^\dagger&=E\hat{G}^\dagger\hat{G}-\hat{G}^\dagger\hat{H}_\mathrm{eff}\hat{G}\\
    \implies \hat{G}&=E\hat{G}^\dagger\hat{G}-\hat{G}^\dagger\hat{H}_\mathrm{eff}^\dagger\hat{G}\\
    \implies \hat{G}^\dagger-\hat{G}&=\hat{G}^\dagger\left(\hat{H}_\mathrm{eff}^\dagger-\hat{H}_\mathrm{eff}\right)\hat{G}\nonumber\\
    &=2\pi\mathrm{i}\hat{G}^\dagger\hat{W}\hat{W}^\dagger\hat{G}.\label{eq:Green_identity}
\end{align}
In the last step we used the Hermiticity of $\hat{H}$.

We assume that the coupling operator $\hat{W}$ is independent of $\theta$ and thus
\begin{equation}
    \partial_\theta\hat{U}=2\pi\mathrm{i}\hat{W}^\dagger\hat{G}\left(-\partial_\theta\hat{H}\right)\hat{G}\hat{W}.\label{eq:dU_Green}
\end{equation}

Using Eqs.\ (\ref{eq:Green_identity}) and (\ref{eq:dU_Green}), it is straightforward to show that
\begin{equation}
    \hat{Q}_\theta=2\pi\hat{W}^\dagger\hat{G}^\dagger\left(-\partial_\theta\hat{H}\right)\hat{G}\hat{W}.\label{eq:QWS_Force_monochrom}
\end{equation}
This equation already allows for the desired interpretation: The operator $\hat{W}$ maps the asymptotic region to the boundary of the scattering system and the Green's operator $\hat{G}$ describes the propagation inside the system. So indeed the QWS operator $\hat{Q}_\theta$ can be interpreted as the ``asymptotic counterpart'' to the local force $-\partial_\theta\hat{H}$. To illustrate this relation even further, it is convenient to write
\begin{equation}
    \ket{\psi_\mathrm{scat}}=\sqrt{\varepsilon}\hat{G}\hat{W}\ket{\psi_\mathrm{in}},\label{eq:PsiScat_PsiIn}
\end{equation}
where $\ket{\psi_\mathrm{in}}$ is the input state in the asymptotic region, $\ket{\psi_\mathrm{scat}}$ is the scattering state in the interior of the system and $\varepsilon$ is an auxiliary quantity with the physical unit of energy with the purpose of cancelling the physical units of $\hat{G}$ ($\mathrm{J}^{-1}$) and $\hat{W}$ ($\mathrm{J}^{1/2}$).

Combining Eqs.\ (\ref{eq:QWS_Force_monochrom}) and (\ref{eq:PsiScat_PsiIn}) yields
\begin{equation}
    \braket{\psi_\mathrm{scat}|\left(-\partial_\theta\hat{H}\right)|\psi_\mathrm{scat}}=\frac{\varepsilon}{2\pi}\braket{\psi_\mathrm{in}|\hat{Q}_\theta|\psi_\mathrm{in}},
\end{equation}
which generalizes the central Eq.\ (2) in \cite{Horodynski2020}. Note that this relation is evaluated at a single energy $E$. So in order to get the total force $\hat{K}_\theta$, we have to integrate over the whole energy spectrum. The auxiliary $\varepsilon$ is conveniently replaced by the infinitesimal measure $\mathrm{d}E$:
\begin{equation}
    \hat{K}_\theta=\frac{1}{2\pi}\int_0^\infty\hat{Q}_\theta\,\mathrm{d}E.
\end{equation}
One might very well question why we replace $\varepsilon$ by $\mathrm{d}E$ without any further numerical factors. In Section \ref{sec:VacForces} (see Eq.\ (\ref{eq:K_theta_vac})), we derived the vacuum contribution to $\hat{K}_\theta$ with an independent calculation, which fixes the prefactor to $\left(2\pi\right)^{-1}$.

\subsection{Minimal variance\label{sec:App_Micromanip_MinVar}}

We want to exploit the degeneracy of the optimal input states for micromanipulation encountered in Section \ref{sec:QuMicroMan}. Amongst the Gaussian input states with all $\nu$ photons in the channel $i_\mathrm{max}$, we want to find the ones that minimize the standard deviation (or likewise, the variance) of $\hat{Q}_\theta$. In Appendix \ref{sec:App_QFI_OptGaussian} (which does \emph{not} build on this subsection), we perform some optimization calculations with this quantity and the main ideas there are also relevant here. We invite the reader to study \ref{sec:App_QFI_OptGaussian} before continuing here, so that we do not have to repeat ourselves unnecessarily.

The main differences to \ref{sec:App_QFI_OptGaussian} are that we now want to \emph{minimize} the variance, and that we are already restricted to a single channel, which is why we drop the associated index $i_\mathrm{max}$ in the following.

For the squeezing angle $\psi$ we can conclude, analogously to the considerations after Eq.\ (\ref{eq:VarQ_diagonal_betappsi}), that the optimal value is given by $\psi=2\arg\left(\beta\right)$, which amounts to amplitude squeezing, see also Fig.\ \ref{fig:Squ_state}. The remaining task is to minimize the expression
\begin{equation}
    \left|\beta\right|^2\mathrm{e}^{-2p}+2\cosh^2\left(p\right)\sinh^2\left(p\right)
\end{equation}
under the constraint
\begin{equation}
    \left|\beta\right|^2+\sinh^2\left(p\right)=\nu.
\end{equation}
(The sign in the exponential differs from Eq.\ (\ref{eq:VarQ_diagonal_betap}) due to the different value for $\psi$.)
Inserting the constraint into the target function yields
\begin{equation}
    \min_{p\in\left[0,\operatorname{arsinh}\left(\sqrt{\nu}\right)\right]}\left(\nu\mathrm{e}^{-2p}+\sinh^2\left(p\right)\left(1+\sinh\left(2p\right)\right)\right).
\end{equation}
This minimization problem has a unique nontrivial yet analytically expressible solution for all $\nu\geq0$, found with the help of Wolfram Mathematica:
\begin{align}
    \left|\beta_\mathrm{opt}\left(\nu\right)\right|&=\sqrt{\nu-\sinh^2\left(p_\mathrm{opt}\left(\nu\right)\right)},\\
    p_\mathrm{opt}\left(\nu\right)&=\frac{1}{2}\ln\left(\frac{1}{2}\left(\sqrt{g\left(\nu\right)}\vphantom{\sqrt{\frac{4\left(1+2\nu\right)}{\sqrt{g\left(\nu\right)}}-g\left(\nu\right)}}+\sqrt{\frac{4\left(1+2\nu\right)}{\sqrt{g\left(\nu\right)}}-g\left(\nu\right)}\right)\right),\\
    g\left(\nu\right)&:=\frac{4}{h\left(\nu\right)}+\frac{h\left(\nu\right)}{3},\\
    h\left(\nu\right)&:=\left(54\left(1+2\nu\right)^2\vphantom{\sqrt{2916\left(1+2\nu\right)^4-1728}}+\sqrt{2916\left(1+2\nu\right)^4-1728}\right)^{1/3}.
\end{align}
The functions $g\left(\nu\right)$ and $h\left(\nu\right)$ are mere mathematical auxiliary functions. A plot of $\left|\beta_\mathrm{opt}\left(\nu\right)\right|$ and $p_\mathrm{opt}\left(\nu\right)$ is given in Fig.\ \ref{fig:OptMM_pbeta}.

\section{Quantum Fisher information\label{sec:App_QFI}}

\subsection{QFI of a Gaussian probe state\label{sec:App_QFI_Gaussian}}

We use the $\mathcal{Q}$ representation introduced at the end of Subsection \ref{subsec:QuWSOp}. From Eqs.\ (\ref{eq:QWS_Qrepr}) and (\ref{eq:QFI_VarQ}) we obtain
\begin{equation}
    F_\theta=4\sum_{i=1}^N\lambda_i^2\left\langle\hat{b}_i^\dagger\hat{b}_i\right\rangle+4\sum_{i,j=1}^N\lambda_i\lambda_j\left(\left\langle\hat{b}_i^\dagger\hat{b}_j^\dagger\hat{b}_i\hat{b}_j\right\rangle-\left\langle\hat{b}_i^\dagger\hat{b}_i\right\rangle\left\langle\hat{b}_j^\dagger\hat{b}_j\right\rangle\right),\label{eq:VarQ_diagonal_bij}
\end{equation}
where $\left\langle\cdot\right\rangle$ denotes the expectation value with respect to the pure Gaussian probe state $\ket{\bm{\beta},\bm{\Xi}}^\mathcal{Q}$.

We start by calculating the expectation values $\langle\hat{b}_i^\dagger\hat{b}_i\rangle$, which are the mean photon numbers $\nu_i$ defined after Eq.\ (\ref{eq:QWS_Qrepr}). These expectation values are the same as the squared norms of the vectors $\hat{b}_i\ket{\bm{\beta},\bm{\Xi}}^\mathcal{Q}$. Using the identities (2.15) and (2.17) from \cite{MaRhodes1990} (note the different sign convention in the definition of the squeezing operator), the polar decomposition $\bm{\Xi}=\mathbf{P}\mathrm{e}^{\mathrm{i}\bm{\Psi}}$ of the squeezing matrix, Eq.\ (\ref{eq:SqCohState_def_Q}), $\hat{b}_i\ket{\mathbf{0}}=0$ and denoting $\ket{\mathbf{e}_i}^\mathcal{Q}:=\hat{b}_i^\dagger\ket{\mathbf{0}}$, we calculate
\begin{equation}
    \hat{b}_i\ket{\bm{\beta},\bm{\Xi}}^{\mathcal{Q}}=\beta_i\hat{D}_b\left(\bm{\beta}\right)\hat{S}_b\left(\bm{\Xi}\right)\ket{\mathbf{0}}-\sum_{j=1}^N\left(\sinh\left(\mathbf{P}\right)\mathrm{e}^{\mathrm{i}\bm{\Psi}}\right)_{ij}\hat{D}_b\left(\bm{\beta}\right)\hat{S}_b\left(\bm{\Xi}\right)\ket{\mathbf{e}_i}^\mathcal{Q}.
\end{equation}
Since both the displacement operator $\hat{D}_b\left(\bm{\beta}\right)$ and the squeezing operator $\hat{S}_b\left(\bm{\Xi}\right)$ are unitary, they transform the orthonormal Fock basis (see Subsection \ref{subsec:Ph_num_probe}) into another orthonormal basis. This means that we can simply read off the coefficients in order to calculate the squared norm of this vector:
\begin{equation}
    \left\Vert\hat{b}_i\ket{\bm{\beta},\bm{\Xi}}^{\mathcal{Q}}\right\Vert^2=\left|\beta_i\right|^2+\sum_{j=1}^N\left|\left(\sinh\left(\mathbf{P}\right)\mathrm{e}^{\mathrm{i}\bm{\Psi}}\right)_{ij}\right|^2.
\end{equation}
The second term can be simplified using the Hermiticity of $\mathbf{P}$ and the unitarity of $\mathrm{e}^{\mathrm{i}\bm{\Psi}}$, leading to
\begin{equation}
    \tensor[^{\mathcal{Q}}]{\braket{\bm{\beta},\bm{\Xi}|\hat{b}_i^\dagger\hat{b}_i|\bm{\beta},\bm{\Xi}}}{^{\mathcal{Q}}}=\left|\beta_i\right|^2+\left(\sinh^2\left(\mathbf{P}\right)\right)_{ii}.\label{eq:Exp_bibi_Gauss}
\end{equation}
Analogously, we get in the $\mathcal{M}$ representation:
\begin{equation}
    \tensor[^{\mathcal{M}}]{\braket{\bm{\alpha},\mathbf{Z}|\hat{a}_m^\dagger\hat{a}_m|\bm{\alpha},\mathbf{Z}}}{^{\mathcal{M}}}=\left|\alpha_m\right|^2+\left(\sinh^2\left(\mathbf{R}\right)\right)_{m,m}.
\end{equation}

The analytical calculation of the second line of Eq.\ (\ref{eq:VarQ_diagonal_bij}) is more extensive, but still feasible involving only fundamental algebra. We skip a detailed derivation and state the result:
\begin{equation}
    F_\theta=4\sum_{i=1}^N\lambda_i^2\nu_i+4\sum_{i,j=1}^N\lambda_i\lambda_j\mu_{ij},\label{eq:QFI_Gaussian}
\end{equation}
where the $\nu_i$ are given by Eq.\ (\ref{eq:Exp_bibi_Gauss}) and
\begin{align}
    \mu_{ij}=&\left|\left(\cosh\left(\mathbf{P}\right)\mathrm{e}^{\mathrm{i}\bm{\Psi}^\top}\sinh\left(\mathbf{P}^\top\right)\right)_{ij}\right|^2+\sum_{i'=1}^N\left|\left(\sinh\left(\mathbf{P}\right)\mathrm{e}^{\mathrm{i}\bm{\Psi}}\right)_{ii'}\left(\mathrm{e}^{-\mathrm{i}\bm{\Psi}}\sinh\left(\mathbf{P}\right)\right)_{i'j}\right|^{2}\nonumber\\
    &-2\Re\left(\beta_i^\ast\beta_j^\ast\left(\cosh\left(\mathbf{P}\right)\mathrm{e}^{\mathrm{i}\bm{\Psi}^\top}\sinh\left(\mathbf{P}^\top\right)\right)_{ij}\right)+2\Re\left(\beta_i^\ast\beta_j\left(\sinh^2\left(\mathbf{P}\right)\right)_{ij}\right).
\end{align}

As a special case, the QFI of the coherent state $\ket{\bm{\beta}}^\mathcal{Q}=\ket{\bm{\alpha}}^\mathcal{M}$ is obtained by setting the squeezing matrix to zero. In this case, $\nu_i=\left|\beta_i\right|^2$ and $\mu_{ij}=0$ and therefore (using the representation conversions given at the end of Subsection \ref{subsec:QuWSOp})
\begin{equation}
    F_\theta=4\bm{\beta}^\dagger\bm{\Lambda}^2\bm{\beta}=4\bm{\alpha}^\dagger\mathbf{Q}_\theta^2\bm{\alpha}.
\end{equation}

\subsection{Optimal monochromatic Gaussian probe state\label{sec:App_QFI_OptGaussian}}

It is numerically verified and shown in \cite{Matsubara2019} that the optimal Gaussian state that maximizes the QFI in Eq.\ (\ref{eq:QFI_Gaussian}) has a squeezing matrix $\bm{\Xi}=\mathbf{P}\mathrm{e}^{\mathrm{i}\bm{\Psi}}$ which is diagonal in the $\mathcal{Q}$ representation. This implies that $\mathbf{P}$ and $\bm{\Psi}$ are diagonal as well. We denote the respective diagonal elements with $p_i\geq0$ and $\psi_i\in\left[0,2\pi\right)$. All matrix functions ($\sinh$, $\cosh$ and $\exp$) operate trivially on diagonal matrices. After some further calculations, we obtain from Eq.\ (\ref{eq:QFI_Gaussian}) the intermediate result
\begin{align}
    F_\theta&=4\sum_{i=1}^N\lambda_i^2\left(\left|\beta_i\right|^2\cosh\left(2p_i\right)+2\cosh^2\left(p_i\right)\sinh^2\left(p_i\right)\right.\nonumber\\
    &\left.\vphantom{\left|\beta_i\right|^2}\hphantom{4\sum_{i=1}^N\lambda_i^2\left(\right.}-2\cosh\left(p_i\right)\sinh\left(p_i\right)\Re\left(\beta_i^{\ast 2}\mathrm{e}^{\mathrm{i}\psi_i}\right)\right).\label{eq:VarQ_diagonal_betappsi}
\end{align}
We now take a closer look at the last term of this expression. The product of the hyperbolic functions is always non-negative, so we have to minimize $\Re\left(\beta_i^{\ast2}\mathrm{e}^{\mathrm{i}\psi_i}\right)$ in order to maximize the overall QFI. Since $\psi_i$ and the phase of $\beta_i$ appear only in this term, we can do the optimization with respect to these variables independently of the other terms. By varying $\psi_i$, the minimal value is $\Re\left(\beta_i^{\ast2}\mathrm{e}^{\mathrm{i}\psi_i}\right)=-\left|\beta_i\right|^2$, which is achieved by the choice $\psi_i=2\arg\left(\beta_i\right)+\pi$. This corresponds to the squeezing angles $\psi_i/2=\arg\left(\beta_i\right)+\pi/2$, which simply means phase squeezing, see also Fig.\ \ref{fig:Squ_state}. Having performed the optimization over the $\psi_i$, we are left with
\begin{equation}
    F_\theta=4\sum_{i=1}^N\lambda_i^2\left(\left|\beta_i\right|^2\mathrm{e}^{2p_i}+2\cosh^2\left(p_i\right)\sinh^2\left(p_i\right)\right).\label{eq:VarQ_diagonal_betap}
\end{equation}
In order to obtain a meaningful maximization problem w.r.t.\ the QFI in the last equation, we have to impose certain restrictions on the parameters $\left|\beta_i\right|$ and $p_i$. The most intuitive constraint is to assume that the mean total photon number is equal to a given value $\nu$ (see Eq.\ (\ref{eq:Exp_bibi_Gauss})):
\begin{equation}
    \sum_{i=1}^N\left(\left|\beta_i\right|^2+\sinh^2\left(p_i\right)\right)=\nu.
\end{equation}
Independently of the precise value $\nu$, we get the following optimal state: All channels are populated by the vacuum state and the whole energy (in form of photons) is used to squeeze the vacuum in the channel $i_{\mathrm{hav}}$ corresponding to the eigenvalue $\lambda_{i_{\mathrm{hav}}}$ of the GWS matrix $\mathbf{Q}_\theta$ with the \textbf{h}ighest \textbf{a}bsolute \textbf{v}alue. The squeezing angle $\psi_{i_{\mathrm{hav}}}=2\arg\left(\beta_{i_{\mathrm{hav}}}\right)+\pi$ is not defined in this case where $\beta_{i_{\mathrm{hav}}}=0$, yet its value is irrelevant (see Eq.\ (\ref{eq:VarQ_diagonal_betappsi})), so we set it to zero without loss of generality. The optimal state parameters are therefore
\begin{align}
    \bm{\beta}_{\mathrm{opt}}&=\mathbf{0},\\
    \bm{\Xi}_{\mathrm{opt}}&=\operatorname{arsinh}\left(\sqrt{\nu}\right)\mathbf{e}_{i_{\mathrm{hav}}}\mathbf{e}_{i_{\mathrm{hav}}}^\top.
\end{align}
The expression $\mathbf{e}_{i_{\mathrm{hav}}}\mathbf{e}_{i_{\mathrm{hav}}}^\top$ represents an $N\times N$ zero matrix with a single one-valued entry in row $i_{\mathrm{hav}}$ and column $i_{\mathrm{hav}}$.

\bibliography{Quantum_WS}

\end{document}